\documentclass[12pt]{article}
\usepackage{amsmath,amssymb}
\usepackage[active]{srcltx}
\usepackage{latexsym,graphicx,epsfig}

\def\be{\begin{equation}}
\def\ee{\end{equation}}
\def\bes{\begin{subequations}}
\def\ees{\end{subequations}}
\def\bea{\begin{eqnarray}}
\def\eea{\end{eqnarray}}
\def\bear{\begin{equation}\begin{array}}

\def\bm{\boldmath}

\def\CP{${\cal C}{\cal P}\;$}

\newcommand{\fr}[2]{\frac{{\displaystyle #1}}{{\displaystyle #2}}}
\newcommand{\fn}[1]{\footnote{{\sl\normalsize #1}}}

\newenvironment{Itemize}{\begin{list}{$\bullet$} %
{\setlength{\topsep}{0.2mm}\setlength{\partopsep}{0.2mm} %
\setlength{\itemsep}{0.2mm}\setlength{\parsep}{0.2mm}}} %
{\end{list}}
\newcounter{enumct}
\newenvironment{Enumerate}{\begin{list}{\arabic{enumct}.} %
{\usecounter{enumct}\setlength{\topsep}{0.2mm} %
\setlength{\partopsep}{0.2mm}\setlength{\itemsep}{0.2mm} %
\setlength{\parsep}{0.2mm}}}{\end{list}}
\newsavebox{\fmbox}

\def\NP{{\em Nucl. Phys.\ }}
\def\PL{{\em Phys. Lett.\ }}
\def\PRL{{\em Phys. Rev. Lett.\ }}
\def\PR{{\em Phys. Rev.\ }}

\def\EJ{{\em Eur. Phys. Journ.\ }}
\def\SYaf{{\em Sov. Yad. Fiz.\ }}

\def\vep{\varepsilon}
\def\vak{\varkappa}

\def\Pom{{\bf I\!P}}
\def\od{{\cal O}}

\def\mb{\boldsymbol}
\def\epe{\mbox{$e^+e^-\,$}}
\def\ggam{\mbox{$\gamma\gamma\,$}}
\def\egam{\mbox{$e\gamma \,$}}

\title {\bm \bf Charge asymmetry in \\
 $e^\pm e^-$, $ep$, \egam, \ggam\ collisions}

\author{I.F. Ginzburg\\
  Sobolev Institute of Mathematics,
  Novosibirsk, 630090 Russia, \\
  \& Perugia, INFN, Italy}

\begin{document}

\maketitle

\begin{abstract}

Study of charge asymmetry within a neutral system $\cal
T$ produced in processes $AB\to{\cal T} X$ can help solve
many problems in particle physics. We collect here some
problems, which have been studied well, as well as
proposals for future activity with a brief discussion of the
main results expected in each case. These are:
\begin{Enumerate}
\item {\bf\bm $\epe\to {\cal T} X$.}\hspace{5mm}
  a) {\bm  ${\cal T}=\pi^+\pi^-$, $X=\gamma$}.
     $\rho f\gamma$, $\phi f\gamma$,... vertices.

  b) {\bm ${\cal T}=\pi^+\pi^-$, $K^+K^-$,  $X=\epe$.} Phases
     of $\pi\pi$ scattering, resonances.

  c) {\bf\bm ${\cal T}=c\bar{c},\; b\bar{b}$, $X=\epe$}.
     Discovery of C--even $c\bar{c}$ and $b\bar{b}$ resonances.

  d) {\bf\bm $e\gamma\to  e\bar{t}t$, ${\cal T}= \bar{t}t$}.
     Study of possible CP violation in $t$-quark physics.

\item {\bf\bm $ep\to e\pi^+\pi^- X$}.
  a) Possible discovery of the odderon.

  b) Measuring the phases of the forward {\bm $\gamma p\to \rho p$}
     and {\bm $\gamma A\to \rho A$} amplitudes.

  c) Study of axial current coupling to the Pomeron
     ({\bm$Zp\to f_2X$}).

\item Violation of the quark--hadron duality.

\item Weighted structure functions in DIS.

\item {\bf\bm $e\gamma \to eW^+W^-$.}
Study of possible strong interaction in the Higgs sector.

\item  Polarization charge asymmetry in {\bf\bm\ggam collisions}

\end{Enumerate}

The greatest part of discussion in the paper is phenomenological
with minimization of model dependent details.

\end{abstract}

\section{Introduction}

We consider the processes $AB\to{\cal T} X$ with production of
some truly neutral particle system $\cal T$, which is well
separated from other reaction products $X$. The charge
asymmetry of the reaction products within the system $\cal T$
--- the difference in the distributions of the produced particles
and antiparticles --- can be used as a powerful tool to study
different problems of particle physics. This charge asymmetry
can be of different origin:
\begin{Itemize}

\item \CP violation in production or decay.

\item Specific charge content of the initial state, for example,
      quark content of proton.

\item Interference of production mechanisms
      leading to the same final subsystem $\cal T$ via
      intermediate states with different C--parity.

\end{Itemize}

The well-known example is the forward--backward asymmetry in
the process $\epe\to\mu^+\mu^-$ near $Z$ resonance. The
vector current (mainly, the photon in the intermediate
state) produces a C-odd system, while the axial current in
$Z$ boson produces a C--even system. Their interference
results in the forward--backward asymmetry of the muons.
Another example is the observed charge asymmetry in the
charm photoproduction in the proton fragmentation region
\cite{E691}, which involves the second and the third
mechanisms.

We consider the third mechanism and (in the end) the first
mechanism. In the most considered cases the discussed
charge asymmetry is determined by the controllable C--parity
of the intermediate state, for example, by the photon with
$C=-1$ or $Z$ with axial $C$--even current. The sign of the
asymmetry is defined by the charge sign of the
colliding particles $A$ and (or) $B$.

$\blacksquare$ Hereafter we denote the momenta of colliding
particles as $p_1$ and $p_2$, $s=(p_1+p_2)^2$, $z$--axis is
directed along the collision axis, transverse components of
the momenta are those orthogonal to both $p_1$ and $p_2$, they
are labeled with bold letters. Let the considered neutral
system $\cal T$ contain particles $a_i$ with momenta $p_{ia}$
and antiparticles $\bar{a}_j$ with momenta $p_{j\bar{a}}$.
{\it The operator of charge conjugation for this system
$\hat{C}_{\cal T}$} acts as:
\be
\hat{C}_{\cal T}
M(p_{ia},\,p_{j\bar{a}})=M(p_{j\bar{a}},\,p_{ia})\,.\label{cpidef}
\ee

$\bullet$ {\bf Two--particle final states.} We consider in
detail the production of two--particle systems ${\cal
T}=P^+P^-$, in the processes $\epe\to \epe{\cal T}$,
$e\gamma\to e{\cal T}$, $\gamma p\to {\cal T}X$, $ep\to
e{\cal T}X$, $\ggam\to {\cal T}X$ with $P=\pi$ or $c$, or
$W$, or $\mu$, or ... For such system $\hat{C}_{\cal T}
M(p_+,p_-)=M(p_-,p_+)$.

Let momenta of $P^\pm$ and the corresponding light--cone
variables $x_\pm$, $y_\pm$\ be
\bes\label{def}
\bear{c}
p_\pm=(\varepsilon_\pm,\,{\mb p}_{\pm\bot},\, p_{\pm
z})\,,\;\; x_\pm=\fr{2p_\pm p_2}{s}=
\fr{\epsilon_\pm+p_{z\pm}}{2E_1}\,,\;\; y_\pm=\fr{2p_\pm
p_1}{s}=\fr{m_P^2+p_{\pm\bot}^2}{sx_\pm}\,;\\[4mm] \mbox{\it
We define }\;\boxed{ k=p_++ p_-\,, \quad r= p_+-p_-}\;,\quad
M=\sqrt{k^2}\,,\;\;
\beta=\sqrt{1-\fr{4m_P^2}{M^2}}\,;\\[4mm]
 x=x_++x_-\,,\;\; y=y_++y_-\,,\;\;
t=(p_1-k)^2=-\fr{{\bf{k}}_\bot^2+M^2(1-x)}{x}\,.
\end{array}
\ee
When considering pion pair production (${\cal T}=
\pi^+\pi^-$), we set $\beta=1$.

Let $x$ axis be directed along vector ${\bf{k}}_\bot$ and
$\psi$ be the angle between $x$ axis and some fixed
axis. Besides, for vector $r$ we define the angles in the
c.m.s. of $\cal T$ system as
\be
r_{c.m.s.}=\beta M(0,
\sin\theta\cos\phi\,,\;\sin\theta\cos\phi\,,\; cos\theta)\,.
\ee\ees

$\bullet$ {\bm \bf The phenomenon of charge asymmetry is the
difference in the distributions of particles and
antiparticles.} {\it It is determined by the part of
differential cross section that changes its sign
with $r^\mu\to -r^\mu$ change.}

Particularly, we describe the {\bf  forward--backward (FB)
asymmetry} by variables
 \bes\label{asymdef}
 \be
\xi=\fr{x_+ -x_-}{x} \;\mbox{ or }\;\eta=\fr{y_+-y_-}{y}\;
\mbox{ or }\;
 K_-=\fr{\xi x-\eta y}{x+y}\equiv
\left.\fr{p_{+z}-p_{-z}}{\varepsilon_++
\varepsilon_-}\right|_{c.m.s.}\,. \label{defbasxi}
\ee
Variables $\xi$ and $\eta$ are useful for description of
the FB asymmetry for the system ${\cal T}$ moving approximately
along the 3--momentum of $p_1$ and $p_2$ respectively, while
$K_-$ describes the entire FB asymmetry. For system ${\cal
T}$ moving along $p_1$ (with $x\gg y$) or $p_2$ (with $y\gg
x$) we have, respectively, $K_-\to \xi$ or $K_-\to-\eta$.
Below we mainly consider the first case.

We also describe the {\bf transverse (T) asymmetry} by
variable
\be
v= \fr{\mb{p}_{+\bot}^2-\mb{p}_{-\bot}^2-K_- {\mb
k}_\bot^2}{M|{\mb k}_\bot|}\equiv\fr{({\boldsymbol\rho}_\bot
{\mb k}_\bot)}{M|{\mb k}_\bot|}\;\; \mbox{ with } \;
\boldsymbol{\rho}_\bot=\mb{r}_\bot-K_- \mb{k}_\bot\,.
\label{defbasv}
 \ee

The charge conjugation operator (\ref{cpidef}) acts on
variables introduced as\\
\centerline{$\hat{C}_P K_-=-K_-$, $\hat{C}_P k=k$ and $\hat{C}_P v=-v$.}

When $|k_z|\gg |\bf{k}_\bot|$, simple relations between
the angles in c.m.s. take place:
\be
\xi=\beta\cos\theta\,,\quad v=\beta\sin\theta\cos\phi\,.
\label{xitheta}
\ee\ees

The phase space element for the produced system is
 \be
d\Gamma=\fr{d^3p_+\,d^3p_-}{\vep_+\vep_-}=dt\,dM^2\,dx\,
\,\fr{2dv\,d\xi}{\sqrt{\beta^2-v^2-\xi^2}}\,d\psi\Rightarrow\;
4 \pi\, dt\,dM^2\,dx
\,\fr{dv\,d\xi}{\sqrt{\beta^2-v^2-\xi^2}}\,. \label{phsp}
 \ee
The latter form is obtained after integration over $\psi$.
\vspace{2mm}

In respect to eq.~(\ref{xitheta}), the study of charge
asymmetry for two-particle states is similar to the well
known partial wave analysis.

$\blacksquare$ The magnitude of the asymmetry related to some
C--odd weight function $w$ ($\hat{C}_{\cal T} w=-w$) is
given by integration over some charge symmetric domain
${\cal D}$ ($\hat{C}_{\cal T} {\cal D}={\cal D}$):
 \bes\label{sig}
 \be
\Delta\sigma_w =\int\limits_{\cal D}\fr{w}{\sqrt{<w^2>}}\,
d\sigma\quad \mbox{with }\;<w^2>=\int\limits_{\cal D}
\fr{w^2d\sigma}{\sigma_B}\,, \quad
\sigma_B=\int\limits_{\cal D} d\sigma\,. \label{sig1}
  \ee
In numerical estimates below we use the step functions
$w=\epsilon(K_-)$ or $\epsilon(v)$ for FB or T asymmetries,
respectively $\left(\mbox{with }\;
\epsilon(x)=\left\{\begin{array}{rcc} 1&\;\mbox{at}\;& x>0\\
-1&\;\mbox{at}\;&x<0\end{array}\right.\mbox{ and }
\epsilon^2(x)=1\right)$ :
 \be
\Delta\sigma_K=\int d\sigma(K_->0)- \int
d\sigma(K_-<0)\,,\;\; \Delta\sigma_v=\int d\sigma(v>0)- \int
d\sigma(v<0)\,. \label{sig2}
 \ee

Value of the given asymmetry is determined by its {\it
Statistical Significance} defined via the numbers of signal
and background events \cite{IIvan}. With the integral
luminosity ${\cal L}$:
\be
 SS = \fr{{\cal L}|\Delta\sigma_w|}{\sqrt{ {\cal
 L}\sigma^B}}\,.
  \label{SB}
 \ee
 \ees
In the calculations we use parameters of particles from
ref.~\cite{PDG}.

$\blacksquare$ Standard C--even contributions disappear
in our signal (as the charge asymmetric part of the cross
section disappears in the background). Therefore, to extract
the charge asymmetry signal from the data, it is not
necessary to know the background with high precision. Only
the ratio of signal to statistical fluctuations of the
background is essential.

\section{\bm $\epe\to \pi^+\pi^-+...$, etc.}
\subsection{\bm $\epe\to \pi^+\pi^-\gamma$.
Radiative return studies} \label{radret}

There are two main mechanisms of the dipion production in
this reaction.\\ $\diamondsuit$ Incident $e^-$ (or $e^+$)
emits a bremsstrahlung photon (initial state radiation,
ISR). Next, this electron collides with the positron,
producing C-odd dipion. \\
 $\diamondsuit$ Incident \epe system transforms to dipion
(typically, $\rho$, $\omega$ or $\phi$). Next, this system
emits a photon, turning the dipion to the C--even state
(final state radiation, FSR).\\ The ratio of cross sections
of these processes is of the order of emitting masses ratio,
$\sigma_{FSR}/\sigma_{ISR}\sim (m_e/m_\pi)\lesssim 0.01$.
Nevertheless, better accuracy is necessary in this problem.

To separate ISR and FSR contributions in the data, it was
noted that the dipions produced in ISR and FSR processes
have opposite C--parity, giving charge asymmetry in the
final state. Measuring this charge asymmetry helps separate
the effects and extract the cross section
$\epe\to\gamma^*\to\pi^+\pi^-$ with high precision
\cite{Meln}. Since FSR effect is small, even rough model of
point--like pions (QED) is considered suitable for the
analysis of the experimental data \cite{DAPHNE} to find the
$\epe\to\pi^+\pi^-$ cross section with high precision ({\it
radiative return method}).

Note that in the analysis of these data only FB asymmetry is
considered. Accounting the T asymmetry can also be
useful (its sign is different for pairs moving along
positron and electron directions).

$\Box$ The photons from ISR are concentrated along the
directions of incident electron or positron within the angle
$\sim 1/\gamma \equiv m/E$. The photons from FSR accompany
produced pions and have roughly uniform angular spread.
Thus, the interference is $\lesssim 1/\gamma$ (with
logarithmic enhancement appearing after the detailed
calculation). Its magnitude is about few percent for
$\sqrt{s}\sim 1$ GeV \cite{DAPHNE}, and it disappears
at higher energies.

$\bullet$ One can use this interference to study poorly
known {\bf\bm vertices $\rho\sigma\gamma$, $\rho f_2\gamma$,
$\phi f_0\gamma$, $\phi f_2\gamma$,...} {\it Detailed study
of effective mass $M$ dependence in the charge asymmetry
effects can help separate effects of these couplings.} These
results can give also more precise values of
$\sigma(\epe\to\gamma^* \to\pi^+\pi^-)$.

\subsection{\bm $\epe\to\epe\pi^+\pi^-$,
phases of $\pi\pi$ scattering, resonances.}\label{eepipip}

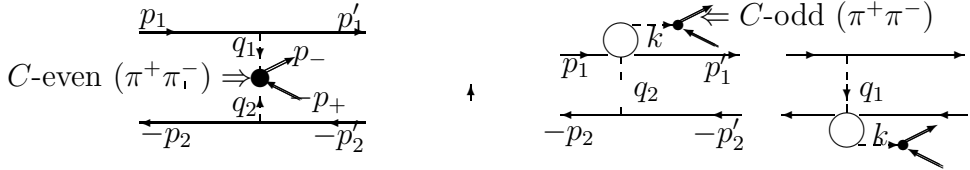
\begin{figure}[htb]
\unitlength=1mm \special{em:linewidth 0.4pt} \linethickness{0.4pt}
\begin{picture}(120.00,20.00)

\put(49.00,15.00){\makebox(0,0)[cc]{$q_1$}}
\put(49.00,7.00){\makebox(0,0)[cc]{$q_2$}}
\put(33,11){\makebox(0,0)[c]{ $C$-even
$(\pi^+\pi^-)\Rightarrow$}}
\put(65.00,18.80){\makebox(0,0)[r]{$p'_1$}}
\put(35.00,18.80){\makebox(0,0)[l]{$p_1$}}
\put(65.00,3.20){\makebox(0,0)[r]{$-p'_2$}}
\put(35.00,3.20){\makebox(0,0)[l]{$-p_2$}}

{\put(51.00,5.0){\line(0,1){1}}
\put(41.00,9.50){\line(0,1){1.00}}
\put(51.00,17.00){\line(0,-1){1}}
\put(51.00,12.50){\line(0,-1){1.00}}

\put(51.00,11.00){\circle*{2.50}}}
\put(50.90,11.25){\line(2,1){4.00}}
\put(50.70,11.45){\line(2,1){4.00}}
\put(52.50,10.15){\line(2,-1){4.00}}
\put(52.30,10.00){\line(2,-1){4.00}}
\put(51.60,10.50){\vector(-2,1){0.00}}
\put(55.70,13.80){\vector(2,1){0.00}}
\put(57.70,13.80){\makebox(0,0)[cc]{$p_-$}}
\put(58.70,8.00){\makebox(0,0)[cc]{$-p_+$}}

\put(51.00,15.00){\vector(0,-1){1.50}}
\put(51.00,7.00){\vector(0,1){1.50}}

\put(35.00,17.00){\vector(1,0){5.00}}
\put(40.00,17.00){\vector(1,0){25.00}}
\put(65.00,5.00){\vector(-1,0){5.00}}
\put(60.00,5.00){\vector(-1,0){25.00}}

\put(115.00,6.00){\vector(-1,0){4.00}}
\put(111.00,6.00){\vector(-1,0){20.00}}

\put(111.30,20.70){\vector(2,1){0.00}}
\put(106.50,18.40){\line(2,1){4.00}}
\put(106.50,18.20){\line(2,1){4.00}}

\put(107.20,17.50){\vector(-2,1){0.00}}
\put(107.90,17.00){\line(2,-1){4.00}}
\put(107.90,17.20){\line(2,-1){4.00}}

\put(100.70,14.00){\vector(1,0){14.00}}
\put(95.00,14.00){\line(1,0){2.20}}

\put(100.50,18.00){\line(1,0){1.00}}
\put(102.50,18.00){\line(1,0){1.00}}
\put(104.50,18.00){\vector(1,0){1.50}}
\put(99.00,16.00){\circle{4.60}}
\put(91.00,14.00){\vector(1,0){4.00}}
{\put(106.6,18){\circle*{1.5}}}
\put(125,19){\makebox(0,0)[c]{$\Leftarrow C$-odd
$(\pi^+\pi^-)$}} \put(79.00,8.00){\vector(0,1){2.00}}
\put(99.00,6.00){\line(0,1){1.00}}
\put(99.00,10.70){\line(0,1){1.0}}
\put(99.00,12.80){\line(0,1){0.8}}
\put(92.00,3.50){\makebox(0,0)[cc]{$-p_2$}}
\put(112.00,3.50){\makebox(0,0)[cc]{$-p'_2$}}
\put(102.40,9.00){\makebox(0,0)[cc]{$q_2$}}
\put(93.00,12.20){\makebox(0,0)[cc]{$p_1$}}
\put(112.00,12.20){\makebox(0,0)[cc]{$p'_1$}}
\put(103.50,16.50){\makebox(0,0)[cc]{$k$}}

\put(145.00,6.00){\vector(-1,0){4.00}}
\put(141.00,6.00){\line(-1,0){10.3}}
\put(127.30,6.00){\vector(-1,0){7.00}}

\put(141.10,4.70){\vector(2,1){0.00}}
\put(136.40,2.40){\line(2,1){4.00}}
\put(136.40,2.20){\line(2,1){4.00}}

\put(137.10,1.50){\vector(-2,1){0.00}}
\put(137.80,1.00){\line(2,-1){4.00}}
\put(137.80,1.20){\line(2,-1){4.00}}

\put(130.70,14.00){\vector(1,0){14.00}}
\put(125.00,14.00){\line(1,0){7.00}}

\put(130.50,2.00){\line(1,0){1.00}}
\put(132.50,2.00){\line(1,0){1.00}}
\put(134.50,2.00){\vector(1,0){1.50}}
\put(129.00,4.00){\circle{4.60}}
\put(121.00,14.00){\vector(1,0){4.00}}
\put(136.5,2){\circle*{1.5}}

\put(129.00,10.00){\vector(0,-1){2.00}}
\put(129.00,6.40){\line(0,1){1.00}}
\put(129.00,10.70){\line(0,1){1.0}}
\put(129.00,12.80){\line(0,1){1.2}}

\put(132.40,9.00){\makebox(0,0)[cc]{$q_1$}}
\put(133.50,3.50){\makebox(0,0)[cc]{$k$}}
\end{picture}
 \caption{ \it Two--photon (left) and bremsstrahlung (right)
production of pion pairs} \label{fig1}
\end{figure}

The charge asymmetry of pions in this process appears due to
interference between the amplitudes given by diagrams at
 fig.~\ref{fig1}.
Open circles in the bremsstrahlung amplitudes describe two
QED diagrams for the virtual Compton scattering. Other
diagrams contribute negligibly to the cross section.

This asymmetry was considered first in Ref.~\cite{CS} for
small values ${\bf k}_\bot^2\ll m_\pi^2$ (which contribute
weakly to the observable effects). The presented equations
from Ref.~\cite{GSS} are free from that limitation (similar
equations were also obtained in Ref.~\cite{Diehl}).

The considered charge asymmetric term in the differential
cross section is the sum of terms given by interference
between two--photon amplitude and the bremsstrahlung
amplitudes with radiation from electron $d\sigma_{2e^-}$ or
positron $d\sigma_{2e^+}$:
\be
d\sigma_{asym}=d\sigma_{2e^-}+d\sigma_{2e^+}\;\mbox{ with }\;\;
\hat{C}_\pi d\sigma_{2e^\pm}=-d\sigma_{2e^\pm}\,.
\ee

The main contribution to the term $d\sigma_{2e^-}$ is given
by almost real photon $q_2$. With the {\em logarithmic
accuracy} (which is about 5\% for modern colliders) this
contribution is given by the convolution of the equivalent
photon  $q_2$ spectrum with the charge asymmetric
interference for the subprocess $e\gamma\to e\pi^+\pi^-$. We
write it in the helicity basis for the subprocess
$\gamma\gamma\to\pi\pi$
\bes\label{eqpipi}
\bear{c}
d\sigma_{2e^-}=dn_\gamma(q_2)\otimes
d\sigma_{e\gamma}\,,\;\;
d\sigma_{e\gamma}=\fr{\alpha^2}{32\pi^3q_1^2 M^2 sx E_1^{'}}
\sum\limits_{a=+,-,0}C^{a+}Re(F^*_\pi M_{a+})\;
d\Gamma_{\pi\pi}\Rightarrow\\[4mm]  d\sigma_{2e^-}=
\fr{\alpha^3}{8\pi^4}\fr{\rho_2^{++}L_2} {s^2x[M^2(1-x)+{\bf
k}_\bot^2]}\;\;\sum\limits_{a=+,-,0}g^{a+}Re(F^*_\pi
M_{a+})\;d\Gamma_{\pi\pi}\,;\\[4mm]
\rho_2^{++}=\fr{2-2y_2+y_2^2}{y_2^2}\;,\;\;\;\;
L_2=\ln\left(\fr{|q_2^2|_{max}(1-y_2)}
{m_e^2y_2^2} \right)\,,\\[4mm]
y_2=\fr{2q_2P_1}{s}=\fr{M^2(1-x)+{\bf
k}_\bot^2}{sx(1-x)}\,,\;\;\; |q_2^2|_{max}\approx
min\left(\fr{{\bf k}_\bot^2}{1-y_2}\,,m_\rho^2\,,M^2\right)
\,.
\end{array}\label{eqpipi1}
\ee
\be
\begin{array}{c}
g^{++}=\xi(2-x) +v \left[(1-x)\fr{M}{|{\bf
k}_\bot|}-\fr{2-2x+x^2}{2(1-x)}\fr{|{\bf
k}_\bot|}{M}\right]\,,\\[4mm]
g^{-+}=-v\left[2-2x+x^2-\fr{{\bf k}_\bot^2}{M^2}\left(\fr{4v^2}
{1-\xi^2}-3\right) \right] \fr{M}{2|{\bf k}_\bot|}
-\xi(2-x)
\fr{2v^2+\xi^2-1}{1-\xi^2}\,,\\[4mm]
g^{0+}=\sqrt{\fr{1-x}{2(1-\xi^2)}}
\left[(2-x)(1-\xi^2)\fr{M}{|{\bf k}_\bot|}-4\xi v
-\fr{2-x}{1-x}(2v^2+\xi^2-1)\fr{|{\bf k}_\bot|}{M}\right]
\,.\end{array}
\ee

The contribution $d\sigma_{2e^+}$ is obtained by changing
the variables
\be
d\sigma_{2e^+}=-d\sigma_{2e^-}(p_1\leftrightarrow
p_2,q_1\leftrightarrow q_2)\,,\;\; M_{ab}(q_1,q_2,\Delta)\to
(-1)^{a+b}M_{ba}(q_2,q_1,\Delta)\,.
\ee
Note that $ \hat{C}_\pi M^{\pm +}= M^{\pm +}\,,\;\;
\hat{C}_\pi M^{0+}=-M^{0+}\,,\;\; \hat{C}_\pi g^{\pm
+}=-g^{\pm +}\,,\;\; \hat{C}_\pi g^{0+} =g^{0+}\,.$ \ees

These equations show that after azimuthal averaging the FB
asymmetry (in $\xi$) does not depend on the value of
amplitude $M_{+-}$, while the T asymmetry (in $v$) includes
both $M_{++}$ and $M_{+-}$ contributions. Comparing this to
the two--photon case, a new term with amplitude $M_{0+}$
appears.

$\Box$ At small $k_\bot$ the considered effects are small,
while the background (mainly two-photon production) is
strongly peaked. At the same time, the effect only weakly
depends on $x$, while two-photon background increases at
$x\approx y$, and bremsstrahlung contribution itself is
strongly peaked at $x\approx 1$ (center and edges of the
rapidity scale). Therefore, cuts on $k_\bot\sim 100$ MeV
(from below) and on $x$ (from both sides) are useful to
increase SS value (\ref{SB}). For the contribution
$d\sigma_{2e^-}$, the transverse momentum of scattered
electron  {\bm $p_{e\bot}=-k_\bot$} with high accuracy.

Here we present some numbers for point--like pions (within
QED).
%
%
%

$\diamondsuit$ For DA$\Phi$NE with $2E=1$ GeV for the
effective mass interval $M=300-350$ MeV with cuts $k_\bot\ge
k_0=100$ MeV and $0.95\ge x,y\ge 0.4$ we have
$\sigma^B=14.6$ pb and $\Delta\sigma_K=-1.07 pb$. At ${\cal
L}=500$~pb$^{-1}$ it gives $SS_K\approx 6.3$. The value of
SS is increased in the mass interval 350-400 MeV.

$\diamondsuit$ For PEPII with $\sqrt{s}=10$ GeV for the
effective mass interval $M=475-525$ MeV with cuts $k_\bot\ge
k_0=150$ MeV and $0.95\ge x,y\ge 0.3$ we have
$\sigma^B=17.2$ pb and $\Delta\sigma_K=-1.62$ pb. At ${\cal
L}=30$ fb$^{-1}$ it gives SS=68!

The strong interaction of pions increases both two--photon
amplitude and (even more) the form--factor as compare QED.
It results in enhancement of SS. Besides, the choice of
suitable cut in $k_\bot$ and weight function should be also
subject of special studies which should enhance SS.

$\blacksquare$ {\bf Physical picture.}

$\diamondsuit$ At $M<500$ MeV the Born QED model (point-like
pions) describes reasonably the $\ggam\to \pi^+\pi^-$
amplitudes, while their phases and the phase of the form
factor reproduce phase shifts for the $S$ and $P$ waves of
the elastic $\pi\pi$ scattering since the unitarity relation
is saturated here with two--pion intermediate states.
Therefore, the data on the charge asymmetry can give us the
energy dependence for the phase shifts $\delta_J^I$ of
$\pi\pi$ scattering via the quantity $\cos(\delta^0_0
-\delta_1^1)$. One can expect to obtain here the precise
values of the scattering lengths. In this energy region
non-trivial effects are given by the first term (with
$M_{++}$) of eq.~(\ref{eqpipi1}) only. For other amplitudes
the point--like QED for pions seems to be a good
approximation.
\begin{figure}[htb]
\label{fig2}
\begin{center}
\includegraphics[width=0.45\textwidth,height=0.16\textheight]
{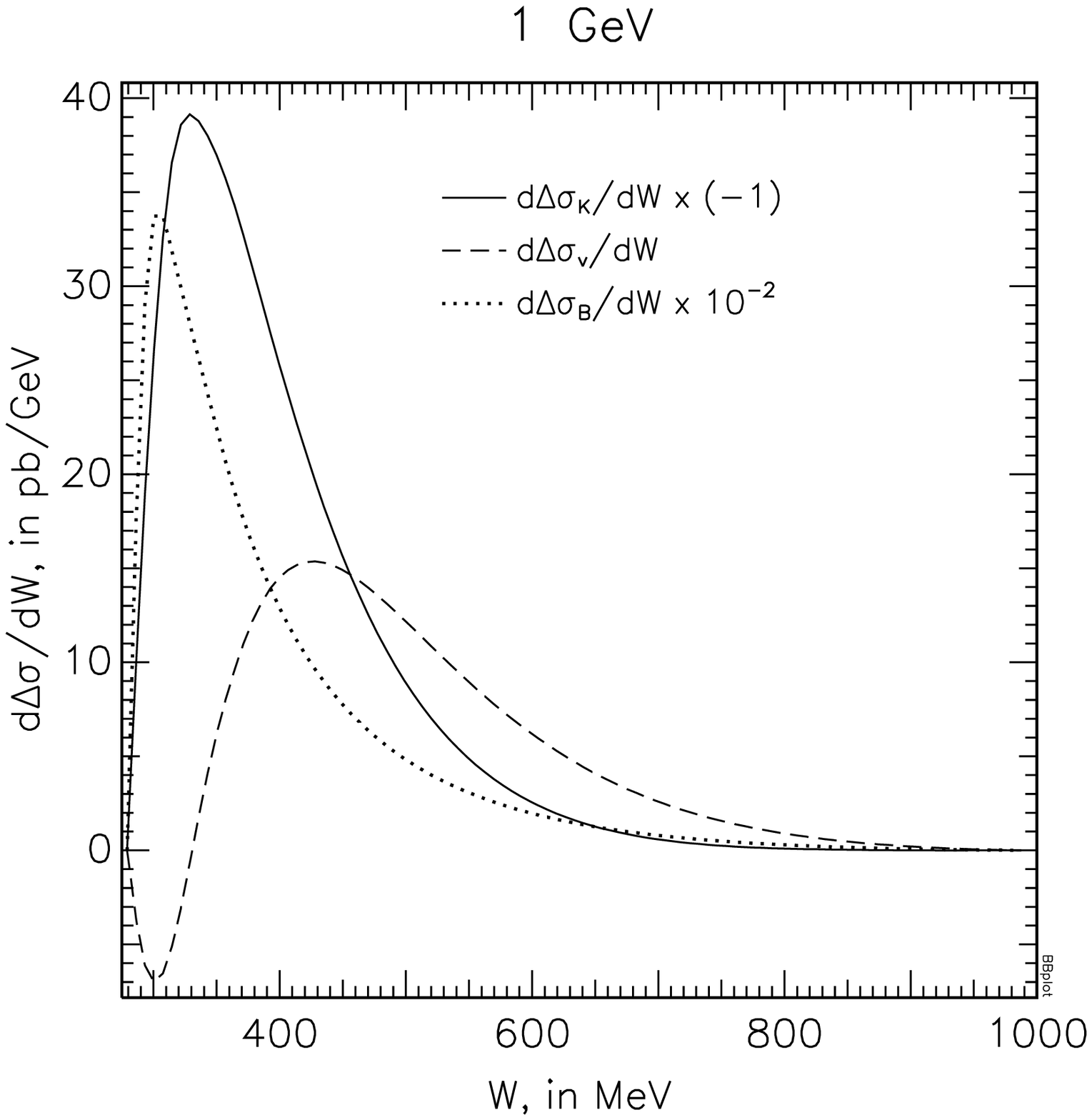}
\includegraphics[width=0.45\textwidth,height=0.16\textheight]
{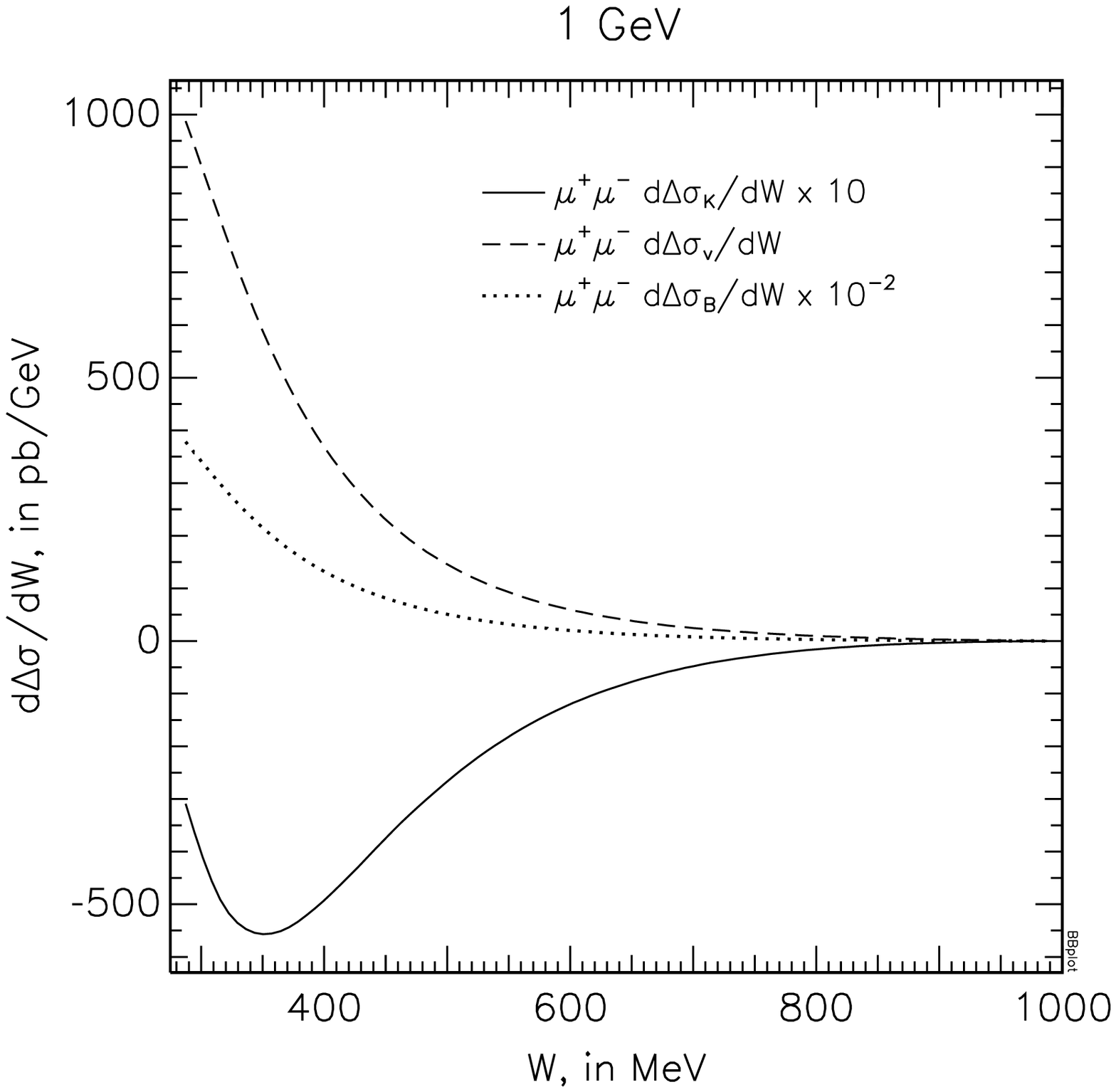}
 \caption{\em The FB ($\Delta\sigma_K$) and T ($\Delta\sigma_v$)
asymmetries and the background, QED for pions -- left panel,
muons -- right panel}
\end{center}
\end{figure}

Let us present some curves for point--like pions at
$\sqrt{s}=1$ GeV \cite{GSS}. The fig.~2 (left panel)
represents integral effects. At low $M$ the
forward-backward asymmetry is higher than the transverse. At
the right panel of fig.~2 we show for comparison the same
asymmetries for muons  in the process $\epe\to
\epe\mu^+\mu^-$ (based on equations from Ref.~\cite{KLMS}).
Comparing contributions
 \begin{figure}[htb]
\label{fig3}\begin{center}
\includegraphics[width=0.45\textwidth,height=0.18\textheight]
{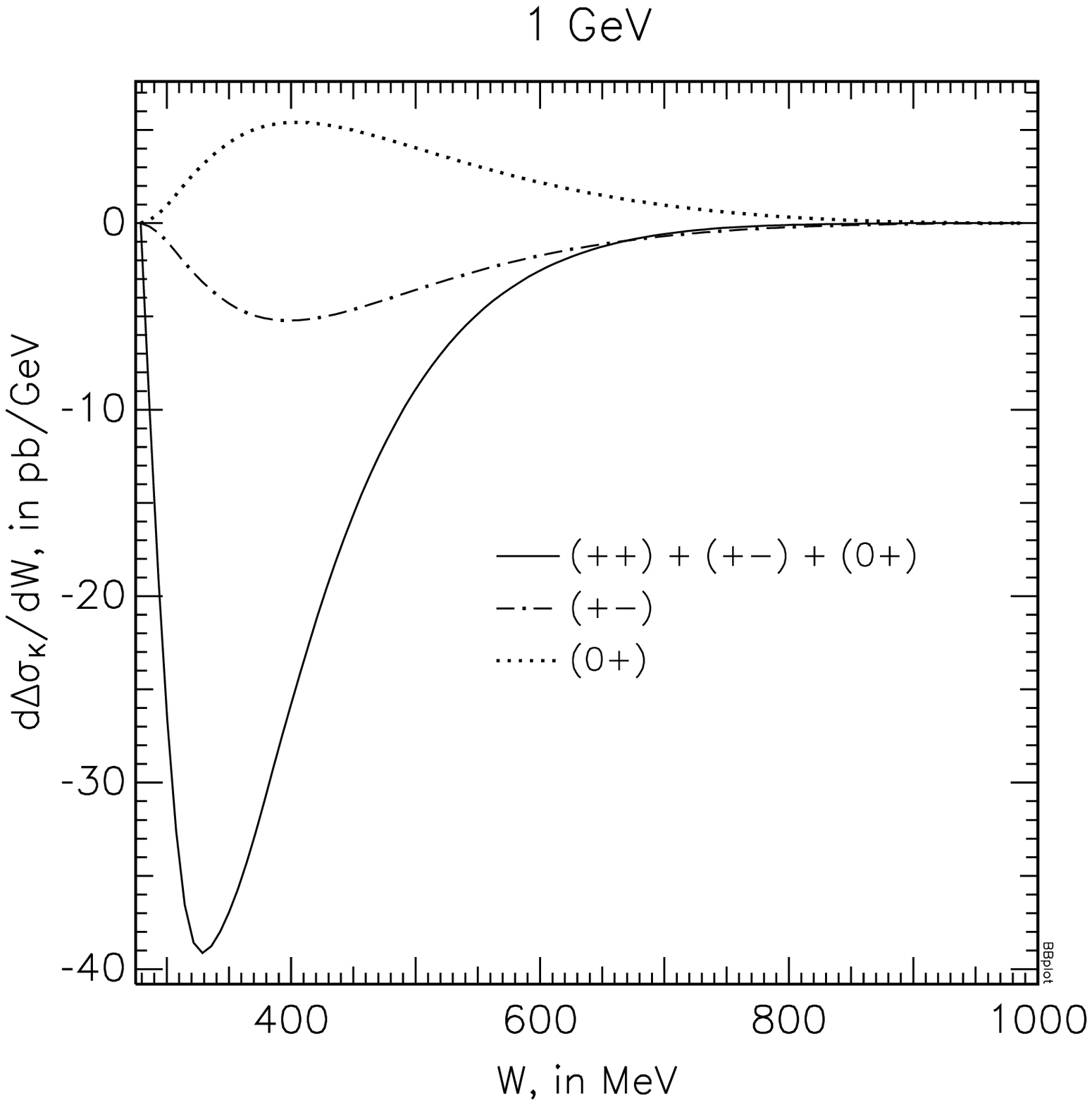}
\includegraphics[width=0.45\textwidth,height=0.18\textheight]
 {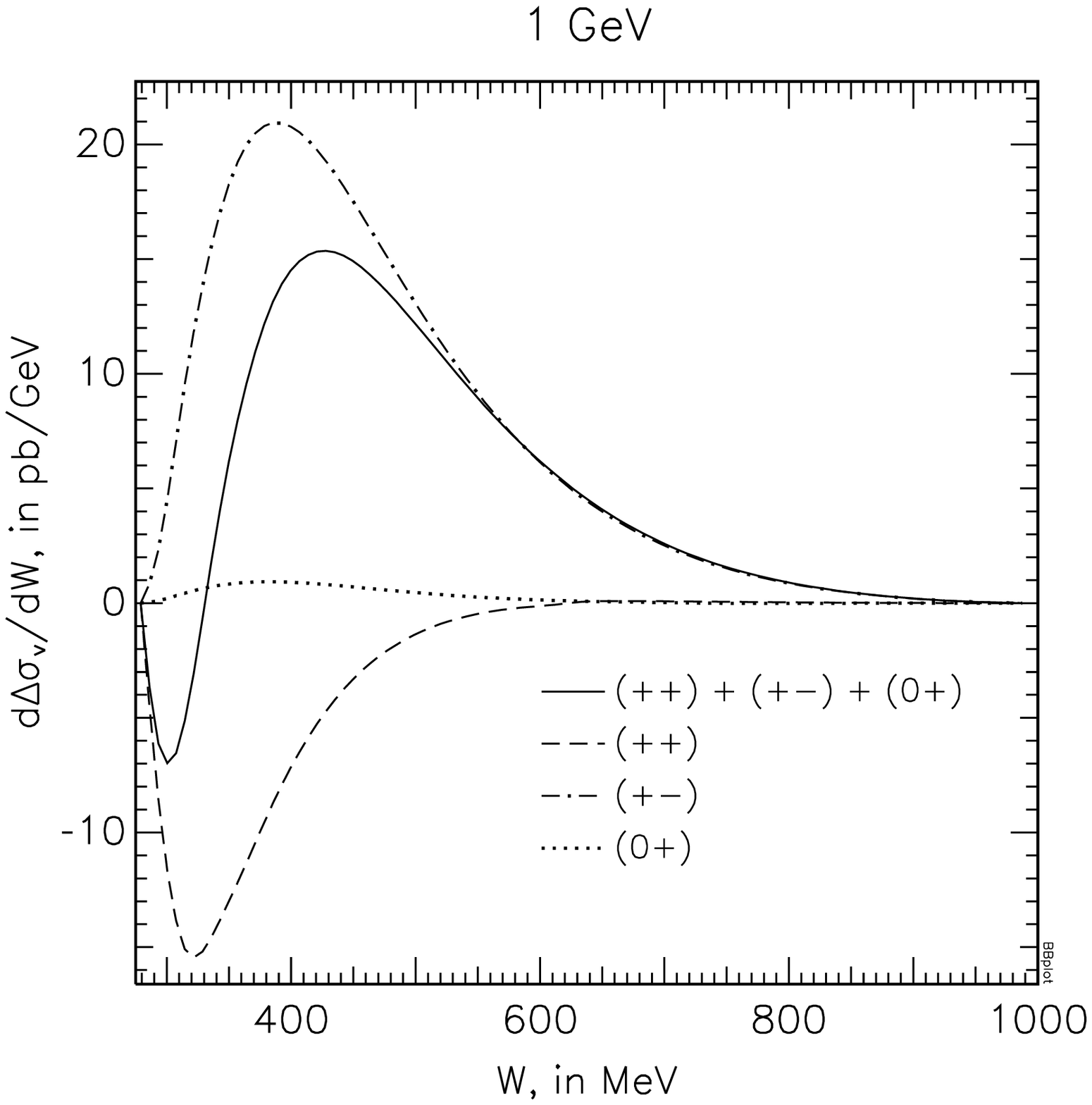}\end{center}
\caption{\em The contributions of different amplitudes to FB
(left panel) and T (right panel) asymmetry }
 \end{figure}
of different helicity amplitudes to FB and T asymmetries
(fig.~3), one can see that in the  FB asymmetry the
amplitude $M_{++}$ dominates. It is useful for the study of
$M\lesssim 1$ GeV and $f_0$'s. In the T asymmetry
contributions of $M_{++}$ and $M_{+-}$ almost compensate
each other. Strong interaction effects generally break
this compensation.

$\diamondsuit$ At higher energies the discussed observations
should help distinguish between different models for the
resonances having two-pion and two-photon decay modes.
For example, it can give us a new information about
$f_0(980)$ and $f_2(1270)$ mesons, etc.
\begin{figure}[htb]
\label{fig4}\begin{center}
\includegraphics[width=0.4\textwidth,height=0.25\textheight]
{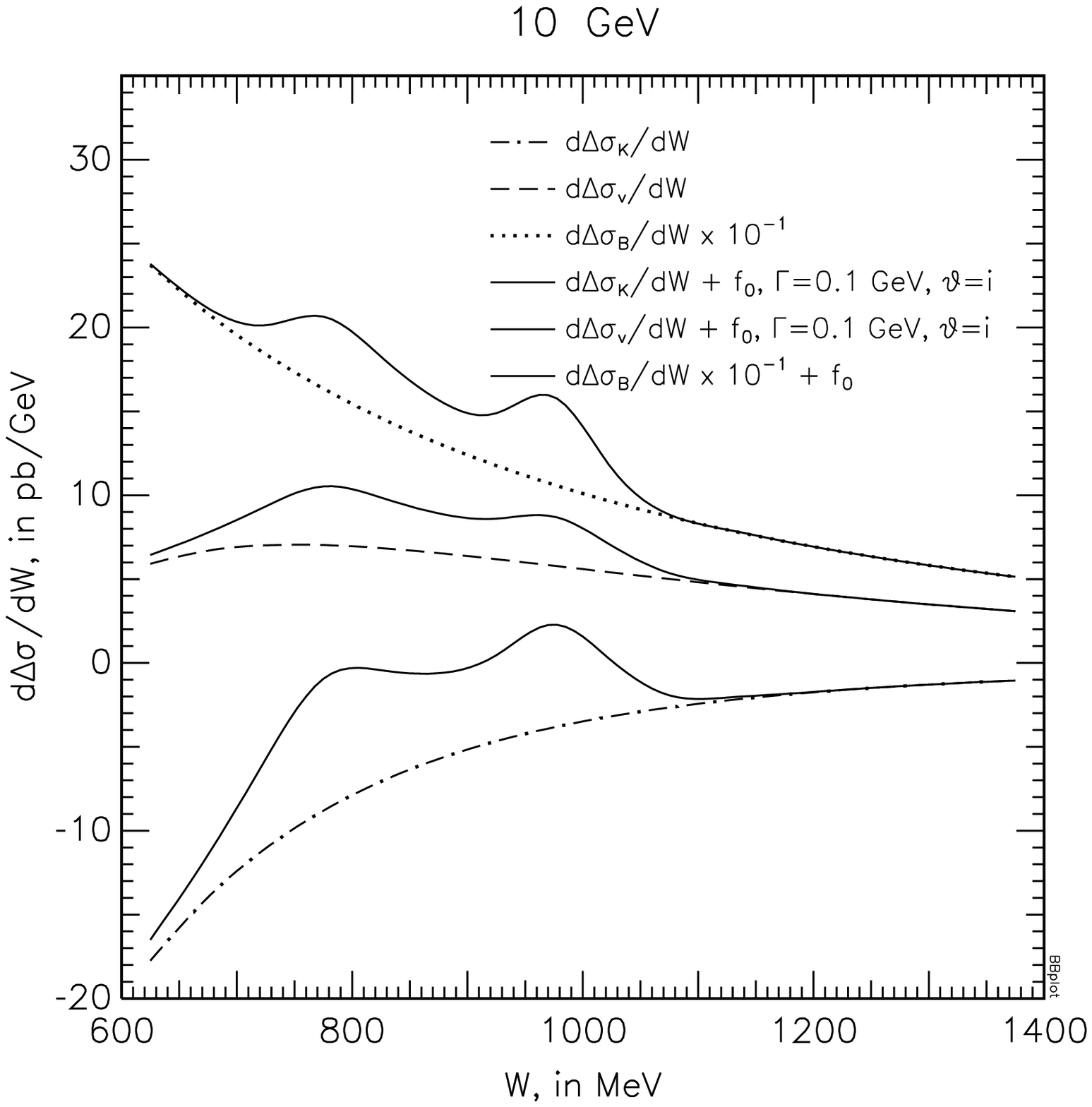}
\includegraphics[width=0.4\textwidth,height=0.25\textheight]
{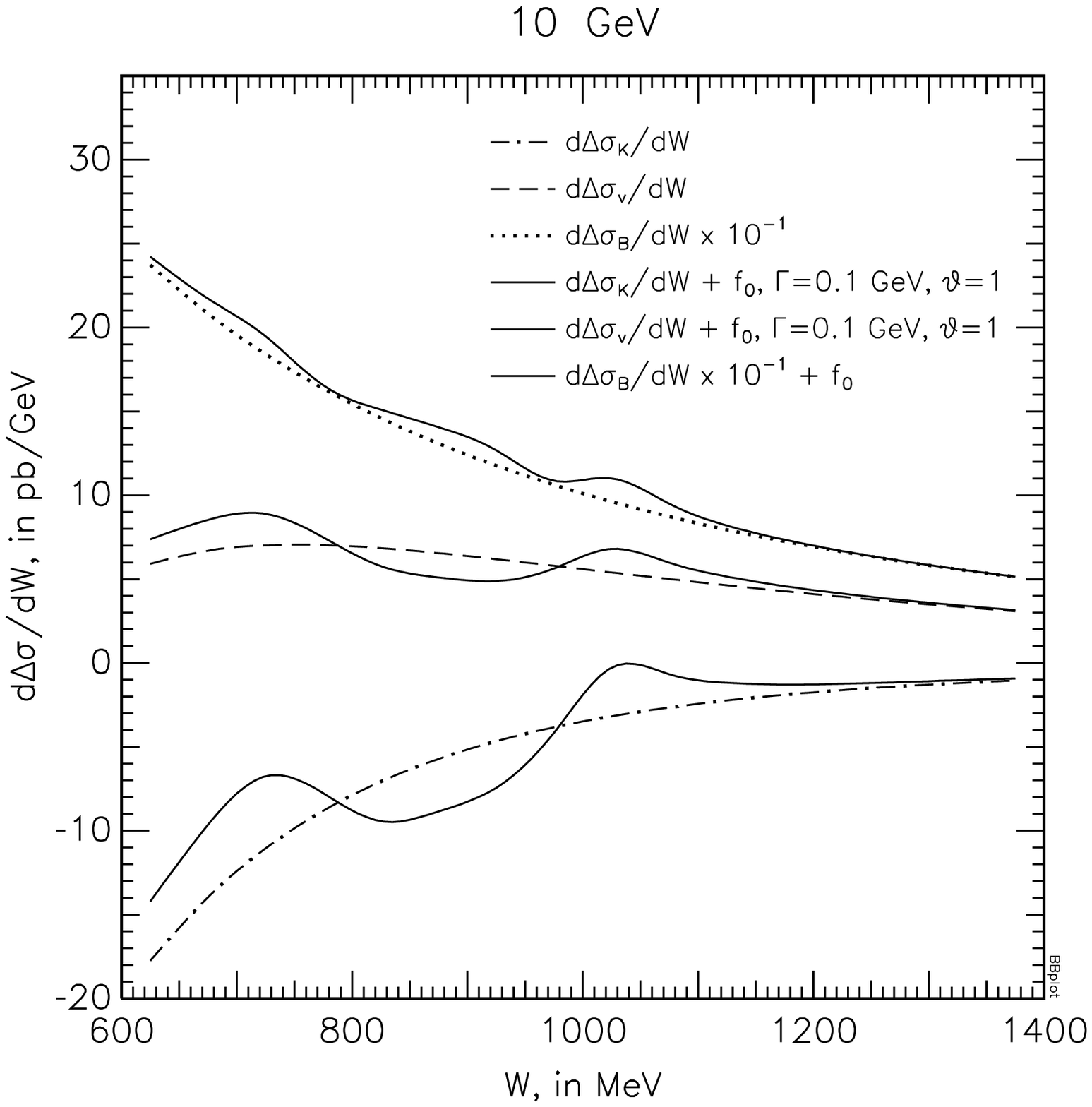}\end{center}
 \caption{\em The charge asymmetries of pions due to
($\rho,\,f_0(980)$) interference.}
\end{figure}

To get an idea about the magnitude of the charge asymmetry
with resonances, we consider a toy model with pion form
factor for $P$--wave + $S$--wave given by ({\it i}) QED +
$f_0(980)$ at $\Gamma_{f^0}= 100$ MeV or ({\it ii}) with
additional phase shift $\vartheta$ (giving effect of
possible $\sigma$ state). In fig.~4 the asymmetries in $K_-$
and $v$ (with suitable kinematical cuts) are compared with
charge symmetric background (solid lines) and pure QED
effects (dotted, etc. lines)

At $M>1.1$ GeV the main non--QED contribution is given by the
term $Re(F_\pi^*M_{++})\propto Re(D_\rho^*D_f)$ ($\rho-f_2$
interference). The {\it overlap factor} which is
proportional to this real part is shown in Fig.~7.

\subsection{Distinguishing processes}

$\bullet$ Two above reactions can be distinguished well, if
in addition to pions, the photon (for $\epe\to
\pi^+\pi^-\gamma$) or scattered electron or positron (for
$\epe\to\epe\pi^+\pi^-$) is observed. For the second
reaction, the major part of the charge asymmetric effect
corresponds to the case when the total transverse momentum
of produced pion pair $\gtrsim 100$ MeV. Therefore, the
scattered $e^-$ (or $e^+$) is recordable almost without loss
of statistics (with scattering angle $\gtrsim 200$ mrad at
DA$\Phi$NE).

$\bullet$ In the inclusive case, with observation of pions
only, these processes can be distinguished with
measurements of the missing mass $M_m$,
 \be
M_m^2=(p_1+p_2-k)^2\,.\label{mm}
 \ee
$\lozenge$ For the process  $\epe\to \pi^+\pi^-\gamma$ it
should be zero ($M_m=0$). To eliminate $\epe\to\pi^+\pi^-\pi^0$
contribution with $\pi^0\to 2\gamma$ decay, one can use cut
$M_m<130$ MeV.\\ $\lozenge$ For the process
$\epe\to\epe\pi^+\pi^-$,  the missing mass is generally
large, $M_m\sim \sqrt{s}$, e.g. one can use cut
$M_m>\sqrt{s}/3$.

\subsection{\bm Kaons, heavy quarks in $\epe\to
\epe{\cal T}$,\ $\egam\to e{\cal T}$}

$\bullet$ Eqs.~(\ref{eqpipi}) also describe charge asymmetry
of {\bf\bm kaons in the process  $\epe\to \epe K^+K^-$}
(with some corrections due to $\beta\neq 1$). At $M_{KK}\approx
1$ GeV, this asymmetry is given by the phase difference
between the amplitudes of $\phi$ meson and still mysterious
$f_0(980)+ a_0(980)$ mesons production.\\
$\bullet$  {\bf\bm Heavy quarks in $\epe\to \epe c\bar{c}$,
$\epe\to \epe b\bar{b}$, $e\gamma\to eb\bar{b}$}. These
heavy quarks should be seen mainly as D (or B) mesons. Near
the threshold these D mesons are produced mainly via
resonance states, excitation of $J/\Psi$ (or $\Upsilon$) for
C-odd states and yet unobserved mesons with spin 0 or 2 for
C--even states. {\bf\bm Charge asymmetry in the
mentioned processes} (e.g., in LEP data) {\bf\bm can help
discover new C--even (with $J=0$ and 2) $c\bar{c}$ and
$b\bar{b}$ resonances.} This problem for the $\egam\to
eb\bar{b}$ process can be considered as the subject for
possible LINX Photon Collider (see \cite{Asner} for details
of project). (The similar analysis of HERA data can also be
useful for such discovery.)\\
$\bullet$  {\bm $e\gamma\to et\bar{t}$}. Effects of New
Physics are expected to be visible well in the interactions
of very heavy $t$--quarks. The analysis of charge asymmetry
in the process will be a new effective tool for the study of
possible \CP violation effects related to New Physics. The
quark-hadron duality works here well since $t$--quark decays
before formation of a bound state. Nevertheless, the
equation for muons,  describing charge asymmetry in QED,
cannot be used here since the picture is strongly changed by
contribution of the axial current from $Z$ boson exchange.
The contribution of the $Z$--bremsstrahlung diagram is
essential in the entire phase space, while the contribution
from the $t$--channel $Z$ boson exchange becomes substantial
with the large transverse momentum of the scattered electron.
These effects should be studied in details to distinguish from
the effects of pure \CP violation related to the New Physics.
The effect of {\em axial $ Z\gamma t\bar{t}$ anomaly} will
also be observed at small transverse momenta of electron
\cite{GIl}.

\section{\bm Diffractive type process $e p\to e\pi^+\pi^-p'$,
${\cal T}=\pi^+\pi ^-$}

In this section we consider the processes in high energy
$ep$ and $eA$ collisions with production of dipion system at
$|t|\ll s$ and separated well from other produced hadrons
(which can be treated as the products of proton excitation
$p'$) --- with large rapidity gap. The discussed processes
are observable at HERA and similar colliders.

To treat the discussed $ep$ collisions with the notation
introduced in Sect.~1, we denote the exchanged photon (or Z)
momentum as $p_1= p_e-p_e^\prime$ and its virtuality
$Q^2=-p_1^2$. Besides, we denote, as usual, with $t (\approx
k_\bot^2)$ the squared momentum transferred from photon (or
Z) to dipion (here ${\bf k}_\bot$ is the transverse momentum
of dipion in the $\gamma^*p$ or $Z^*p$ c.m.s.). The different
values of virtuality $Q^2$ and dipion transverse momentum
$k_\bot$ provide tools to study quite different problems of
the hadron physics.

$\blacksquare$ {\large\bf The mechanisms of dipion
diffractive production. Some features of the case with
almost real photons.}

$\bullet$  The main source of the {\bf\bm C--odd dipions
(mainly $\rho$--mesons}) is the standard Pomeron exchange
($\Pom$) with proton. This amplitude is known well.

$\diamondsuit$ The C-odd dipion can be also produced via
bremsstrahlung mechanism, like in \epe\ collisions. However,
the amplitude of this production is about $\alpha\sim
10^{-2}$ times less than that via Pomeron. It makes this
mechanism negligible and corresponding calculations of
ref.~\cite{kur} meaningless.

$\bullet$ The  mechanisms of {\bf\bm the C-even dipion
production} (e.g., in $f_2$(1270) state) are enumerated in
the table,
 \be \mbox{
\begin{tabular}{|p{2cm}||p{2.2cm}|p{1.8cm}|p{3cm}|p{2.5cm}|}\hline
mechanism&reggeon $\rho$, $\omega$&Primakoff& odderon&
$Z^*\Pom$ collision\\\hline
 where substantial&  $\;\sqrt{s}_{\gamma^*p}\lesssim\;$
$\;\;\;\;\;\;\;$  10~GeV&
 $|k_\bot|<$ \;\; 100~MeV&
$|k_\bot|>200$ MeV, small $Q^2$& $\;\;Q^2\gtrsim\;$
1000~GeV$^2$\\\hline\end{tabular}}
 \label{mechanisms}\ee

$\diamondsuit$ {\it The reggeon  $\rho,\;\omega$ exchanges}.
The corresponding cross section is obtained by simple Regge
extrapolation from low energy data. In the main HERA energy
range this cross section is\ \  $\sim 0.15-0.3$ nb which is
extremely small for observation.

$\diamondsuit$ {\it The Primakoff effect} (dipion production
in collision of incident photon with photon emitted by the
proton) is discussed in detail in sect.~\ref{secpom}. Its
cross section is about 8 nb for $f_2$ production. It is
concentrated within a narrow interval of $k_\bot$ while at
$k_\bot\gtrsim 300$ MeV it is about 0.2 nb, i.e. is
negligible.

$\diamondsuit$ {\it The odderon exchange} is of great
interest for the hadron physics. The odderon is yet elusive
but necessary element of the QCD motivated hadron physics.
The modern status of the odderon in comparison with the
Pomeron is discussed in Appendix. It looks reasonable to
expect that this contribution is not very low and its
transverse momentum dependence is similar to that for other
reggeons.

$\diamondsuit$ At large virtuality $Q^2$ in addition to the
photoproduction of dipions by virtual photon, one should
consider also the $Z$-production of dipions, where virtual
$Z$ is emitted from the electron. Via the dominant Pomeron
exchange, vector part of $Z$-boson produces C--odd dipions
like photon, while its axial part produces C--even dipions.

$\blacksquare$ In sect. 3.1-3.3 we consider {\it processes
with almost real photons} ($Q^2\approx 0$ --- small or
negligible electron escape angle) treating them as $\gamma
p\to \pi^+\pi^-p'$ processes. For these studies, {\it
recording of scattered electron and proton looks
unnecessary}. Therefore, in the estimates of the statistical
significance SS (\ref{SB}) for HERA experiments we use
integrated effective luminosity and the $\gamma p$ center of
mass energy
 \be
{\cal L}_{\gamma p}\approx 100 \mbox{ nb}^{-1}\;\; \mbox{
for }\qquad\sqrt{s_{\gamma p}}\approx 100\div 200\mbox{
GeV}.\label{lum}
 \ee
Recalculations for other values of luminosity and $\sigma_f$
are evident.

\subsection{\bm $Q^2\approx 0$, $k_\bot\gtrsim 200$ MeV:
possible discovery of the odderon} \label{secodd}

The goal of the discussion in this section (based on ref.~
\cite{GIN1}) is to show that the odderon effect can be
discovered via discussed charge asymmetry. In this respect
we prefer to use phenomenological estimates as wide as
possible.

The C--odd dipions are produced via Pomeron exchange, and at
$k_\bot\ge 200$ MeV at HERA only the odderon exchange can
produce C--even dipions with not diminishing rate. The
interference between these amplitudes results in charge
asymmetry of produced pions. In ref.~\cite{GIN1} we propose
to record this very charge asymmetry at HERA for the {\bf
discovery of the odderon}.

$\blacksquare$ {\bf Amplitudes}. Assuming that the Pomeron
$\Pom$ and the odderon $\od$ are Regge poles, their
contributions to the scattering amplitude $AB\to CD$ have
the standard form
 \be
A_{\cal H}=\zeta({\cal H})e^{i\pi \alpha_{\cal H}/2}G(A{\cal H}C)
s^{\alpha_{\cal H}}G(B{\cal H}D)\; \mbox{ with }\;{\cal
H}=\Pom,\,\od\,,\;\;\zeta(\Pom)=1\,,\;\zeta(\od)=i\,.\label{basform}
 \ee
Here the factors $G(A{\cal H}C)$ and $G(B{\cal H}D)$
describe couplings $A{\cal H} C$ and $B{\cal H}D$
respectively. Additional factor $i$ in the odderon amplitude is
related to the opposite signature of the odderon as compared
to the Pomeron. At small $|t|\approx k_\bot^2$, the
dependence of reggeon amplitudes on the difference of
helicities in vertexes $G(A{\cal H}C)$ and $G(B{\cal H}D)$
is given by factors
 \be
G(A{\cal H}C)\propto |t|^{|\lambda_A-\lambda_C|/2}\,,\quad
G(B{\cal R}D)\propto|t|^{|\lambda_B-\lambda_D|/2}
\,.\label{bashel}
 \ee

Let us summarize some general features of these amplitudes
for the diffractive $\gamma p\to \pi^+\pi^- p'$ process
($A=\gamma\,, \,B=p\,,\,C=\pi^+\pi^-,\,D=p^\prime$). In our
discussions we have in mind that $\alpha_\od\sim \alpha_\Pom
\sim 1$ and $\alpha_\Pom-\alpha_\od\ll 1$.

$\bullet$ The {\bf Pomeron amplitude} is studied well at HERA.

1. The main contribution to the cross section is given by
amplitudes with production of two pions in the C-odd state
($\rho$--meson + other $\rho$ type resonances at  higher
effective masses) -- ({\it $\gamma\Pom \rho$} vertex).
Besides, the $s$--channel helicity conservation (SCHC) takes
place at small $t$ --- i.e., $\rho$--meson helicity coincides
with that of the initial photon.

2. The  {\it vertex $p \Pom p'$} is the most significant
when $p'$ coincides with proton $p$ (the admixture from
proton dissociation to excited states with masses $M'
\lesssim 2$ GeV is below 25 \% ). SCHC takes place
for this vertex with good accuracy, $\Delta\lambda_p=0$.

$\bullet$ For the {\bf odderon amplitude} we can use
theoretical estimates only.

1. {\it The vertex $\gamma\od \pi^+\pi^-$} is of main
interest to us. We assume that --- as it is customary for
other phenomena at $M\lesssim 1.5$ GeV --- the pion pairs
are produced mainly via resonance states ($f_0$ and $f_2$
mesons). At $M\gtrsim 1.1$ GeV, we deal here with {\it the
$\gamma\od f_2(1270)$ vertex}. below we consider several
variants of its helicity structure.

2. {\it The vertex $p\od p'$}. In the reggeized 3-gluon
exchange quark--diquark model \cite{Zakharov} (which is
also used --- in some variant --- in ref.~\cite{dosh}) at
small $t$ the properties of the $p\od p'$ vertex are similar
roughly to those of $p\Pom p'$ vertex (see Appendix).
Therefore, we assume the amplitudes with $p'=p$ and SCHC in
this vertex to be either dominant or contributing not less
than other amplitudes.

$\bullet$ The conventional approximation for the amplitudes
of dipion production in the state with angular momentum $J$
and helicity $\lambda$ (with SCHC in proton vertex) is
obtained from eq.~(\ref{basform}) by adding factors
$D_J(M^2)$ and ${\cal E}^{J,\lambda}_{\lambda_\gamma}$ which
describe mass and angular dependencies for decay
$R\to\pi^+\pi^-$ respectively:
  \bes\label{ampl}\be
{\cal A}_{\cal H}=A^{\lambda\lambda_{\gamma}}_{\cal H}\,
D_J(M^2)\;{\cal E}^{J,\lambda}_{\lambda_{\gamma}}\;\;\;
\mbox{ with }\;{\cal H}=\Pom,\,\od\,. \label{basampdef}
 \ee

In our resonant approximation we write $D_J\to D_R$ with $R$
being one of the enumerated resonances. Some of our final
equations are written for the region 1.1 GeV$<M<1.5$ GeV
where C--even dipions are produced in the $f_2$ meson state
while the production of C-odd dipion is described by the
$\rho$ meson tail. The $\rho^\prime$ contribution for the
Pomeron amplitude can easily be implemented in our
equations. At $M\lesssim 1.1$ GeV the $J=1$ and $J=0$
interference can also easily be described with the equations
written below.

$\diamondsuit$ Taking (\ref{bashel}) into account, we specify
the first factor, describing the Regge amplitude of
production of the resonance $R$ with helicity $\lambda_R$,
as
 \be
A^{\lambda_R \lambda_{\gamma}}_{\cal H} = \zeta({\cal H})\;
g^\lambda_R\;\sqrt{\sigma_R B_R}\;\;e^{i\pi\alpha_{\cal
H}/2}\;\; e^{-B_R|t|/2}\; \fr{\left(
B_R|t|\right)^{|\lambda_\gamma-\lambda_R|/2}}
{\sqrt{|\lambda_\gamma-\lambda_R|!}} \,.\label{resampl}
 \ee
The quantity $|g^\lambda_R|^2$ is the fraction of total
cross section of the production of resonance $R$ with
helicity $\lambda_R$, determined by the initial photon with
helicity 1. Due to $P$--invariance, this very quantity is
related to the transitions of photon with helicity $-1$ to
dipion with helicity $-\lambda_R$. The SCHC for Pomeron in
the photon vertex means that $g^1_\rho\approx 1\gg
|g^0_\rho|$. Below we neglect the dependence of parameters
on $M$ (the energy dependence is included in the quantity
$\sigma_R$.)

$\lozenge$ The second factor in eq.~(\ref{basampdef})
describes the dependence of the production amplitude for the
dipion state with spin $J$ on $\pi^+\pi^-$ invariant mass.
We use the standard Breit-Wigner propagation of a resonance
with its coupling to pions even at $|M^2-M_R^2|>
M_R\Gamma_R$,
 \be
D_R(M^2) = \fr{\sqrt{m_R\Gamma_R Br(R\to\pi^+\pi^-)}}{\sqrt{\pi}(
M^2 -m_R^2 + im_R\Gamma_R)}\,.\label{BW}
 \ee

$\lozenge$ The decay factor ${\cal E}^{J
\lambda_{R}}_{\lambda_{\gamma}}$ describes the angular part
of the helicity amplitude. Because pions are spinless, it is
expressed via the standard angular momentum wave functions
$Y_{lm}(\theta,\phi)$ as ${\cal E}^{J,
\lambda_{R}}_{\lambda_{\gamma}} =
Y_{J\lambda_R}(\theta,\phi) e^{i\lambda_{\gamma}\psi}$.\ees
\vspace{2mm}

$\blacksquare$ The charge asymmetry effect is given by the
interference of the Pomeron and the odderon amplitudes
integrated over the redundant phase space variables
(\ref{phsp})
 \be
d\sigma_{asym} = \sum\limits_{\lambda_f,\lambda_\rho}
2Re\left({\cal A}^{\lambda_\rho\dagger}_\Pom {\cal A}
^{\lambda_f}_O\right)d\Gamma\,. \label{dasym}
 \ee

\begin{figure}[!htb]
  \centering
  \epsfig{file=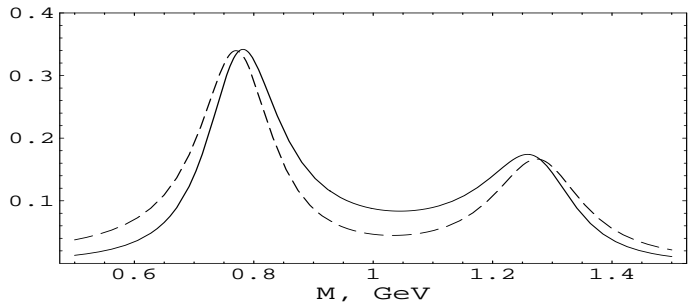,width=10cm,height=4cm}
  \caption{The $\rho-f_2$ overlap function ${\cal I}_{12}(M^2)$
           calculated for $\alpha_\Pom -\alpha_O = 0$ (solid
           line) and $\alpha_\Pom -\alpha_O = 0.2$ (dashed line).}
  \label{figov}
\end{figure}

$\Box$ Let us consider the interference of the $\rho$ meson
production with the odderon mediated $f_2$ meson production
($M>1.1$ GeV). In our approximation its $M$--dependence is
given by the helicity-independent {\it overlap function},
related to the difference between Pomeron and odderon
intercepts $\delta_{\Pom \od} = (\pi/2) (\alpha_{\Pom}
-\alpha_\od)$ as
 \bear{c}
{\cal I}_{\rho f}(M^2) = Re\left[D_\rho(i D_f)^\dagger
 e^{i\delta_{\Pom\od}}\right]\\[4mm]
= Im\left(\fr{e^{i\delta_{\Pom O}}\sqrt{m_\rho
m_f\Gamma_\rho\Gamma_f
Br(f_2\to\pi^+\pi^-)Br(\rho\to\pi^+\pi^-)}}
{\pi(M^2-m^2_\rho+ i m_\rho\Gamma_\rho) (M^2-m_f^2-i m_f
\Gamma_f)}\right)\,.
 \end{array}\label{i}
 \ee
This overlap function, shown in Fig.~\ref{figov}, depends on
the phase difference $\delta_{\Pom O}$ only weakly.

If the difference between Pomeron and odderon intercepts is
small, the overlap function is large ($\sim 1$) when the
phase shift between two Breit-Wigner factors is close to
$\pi/2$. This happens in a wide enough region around the
resonance peaks, where the $D_{R_1}$ (for one resonance) is
almost real while the $D_{R_2}$ (for the other one) is
almost imaginary.

$\Box$ We consider the cross sections averaged over electron
scattering angle, i.e. over initial photon spin states.
Integration over $\psi$ leaves in the result only terms with
identical $\lambda_\gamma$. Besides, due to P--invariance,
for real photons ($\lambda_\gamma=\pm 1$) the other factors
in eq.~(\ref{basampdef}) depend only on the helicity flip
$|\lambda_R-\lambda_\gamma|$, not on the value of helicity
itself. Therefore, the interference effects become
proportional to sums over opposite initial photon helicities
with simultaneous change of sign of final dipion helicities
 \be
{\cal E}^{*J,\lambda_{\rho}}_{\lambda_{\gamma}}{\cal
E}^{J,\lambda_{R}}_{\lambda_{\gamma}}+{\cal
E}^{*J,-\lambda_{\rho}}_{-\lambda_{\gamma}}{\cal
E}^{J,-\lambda_{R}}_{-\lambda_{\gamma}}\propto
\cos[(\lambda_\rho-\lambda_R)\phi] \,.\label{angaver}
 \ee
Since $|J_R-J_\rho|$ is odd, this quantity changes sign with
$\theta\to\pi-\theta$, $\phi\to \pi+\phi$ (i.e.
$p_-\leftrightarrow p_+$). In particular,\\ {\bm \bf The
terms with odd $\lambda_\rho- \lambda_R$} change sign with
$\phi \to\pi+\phi$, {\em i.e.} with $v\to -v$. They are
{\bf responsible for the T asymmetry}.\\ {\bf\bm The terms
with even $\lambda_\rho-\lambda_R$} remain invariant under
$\phi \to\pi+\phi$. Therefore, they must change sign with
$\theta\to \pi-\theta$, i.e. they {\bf are responsible for
the FB asymmetry.}

$\blacksquare$ Neglecting contributions with higher helicity
flips $|\lambda_R-\lambda_\gamma|>1$ and taking into account
explicit forms for spherical harmonics, we obtain final
interference (C--odd) contribution to the cross section in
the form
 \bear{c}
\fr{d\sigma^{interf}}{dM^{2}\,
d\Delta^{2}\,d\xi\,dv}=\fr{3\sqrt{5}{\cal I}_{\rho f}(M^2)}
{2\pi\sqrt{1-\xi^2-v^2}} \sqrt{\sigma_\rho \sigma_f B_\rho
B_f} \exp\left(-\fr{B_\rho + B_f}{2}|t|\right)\otimes
T\,;\\[4mm]
 T= g_\rho^1g_f^1 (1-\xi^2)\xi+\sqrt{|t|}\left\{vg_\rho^1\left[\fr{1}{2}g_f^2(1-\xi^2)+
\fr{1}{\sqrt{6}}
g_f^0(3\xi^2-1)\right]\sqrt{B_f}\right.\\[4mm] \left.
+g_\rho^0g_f^1v\xi^2\sqrt{B_\rho}+
 \xi g_\rho^0\left[\fr{1}{\sqrt{2}}g_f^2(2v^2+\xi^2-1)+
 \fr{1}{\sqrt{3}}g_f^0(3\xi^2-1)\right]\sqrt{B_fB_\rho|t|}\right\}\,.
 \end{array}\label{mainasym}
 \ee

$\bullet$ {\bf The forward--backward asymmetry} is obtained
from here by integration over $v$:
 \bear{c}
\fr{d\sigma_{FB}}{dM^{2}\,
d\vec{k}^{2}\,d\xi}=\fr{3\sqrt{5}}{2} {\cal I}_{\rho f}(M^2)
\sqrt{\sigma_\rho \sigma_f B_\rho B_f}\exp\left(-\fr{B_\rho
+ B_f}{ 2}|t|\right)\otimes \xi T_\xi\,,\\[4mm]
 T_\xi=g_\rho^1g_f^1(1-\xi^2)+
\fr{1}{\sqrt{3}}g_\rho^0g_f^0(3\xi^2-1)\sqrt{B_fB_\rho}|t|\,.
\end{array}\label{xiasym}
\ee

The first term is dominant at small $t$. If the SCHC holds
for the odderon, then the principal effect would be the FB
asymmetry dominated by this first term. If the mechanism of
the $f_2$ production significantly violates SCHC, then the
first term is dominant only at small $t$. With the growth of
$|t|$, the terms with helicity flip both for the Pomeron and
odderon become essential, and generally, not small. Note
that upon the azimuthal integration over entire region of
$v$ variation, the contribution from production of $f_2$ in
the state with helicity 2 vanishes because $\int \cos2\phi
d\phi=0$.

$\bullet$ {\bf The transverse asymmetry} is obtained from
(\ref{mainasym}) by integration over $\xi$:
 \bear{c}
\fr{d\sigma_T}{dM^{2}\, d\vec{k}^2\,dv}=\fr{3\sqrt{10}}{4}
{\cal I}_{\rho f}(M^2) \sqrt{\sigma_\rho \sigma_f B_\rho
B_f}\exp\left(-\fr{B_\rho +
B_f}{2}\right)\otimes\sqrt{|t|}\; vT_v \,,\\[4mm]
  T_v=g_\rho^1 g_f^2 \sqrt{B_f}\;\fr{1+v^2}{2\sqrt{2}}
 +g_\rho^1 g_f^0\sqrt{B_f} \;\fr{1-3v^2}{2\sqrt{3}} +
 g_\rho^0 g_f^1\sqrt{B_\rho}(1-v^2)\,.
 \end{array}\label{vasym}
 \ee
This asymmetry is dominant in the case of strong s--channel
helicity nonconservation (SCHNS) for odderon, for instance,
if the $f_2$ meson is produced in the state with maximal
helicity $\lambda_f =\pm 2$.

The T asymmetry (\ref{vasym}) becomes naturally small at
small $t$ where background is high. Therefore, imposing cut
from below in $|t|$ improves the signal to background ratio.

$\bullet$ The main background to the discussed
Pomeron--odderon charge asymmetry is given by the
Pomeron--photon (Pomeron--Primakoff) interference which is
predominantly transverse (Primakoff mechanism produces
$f_2$ only in the states with helicity 2 or 0). To suppress
this background we introduced cuts in $k_\bot$ which are
different for the FB and T asymmetries:
 \bear{c}
 |t_{FB}|=\vec{k}_\bot^2\ge 0.1B_\rho^{-1}\approx 0.01\,\mbox{
 GeV}^2\Rightarrow k_{\bot FB}>100\mbox{ MeV}\,;\\[3mm]
|t_T|=\vec{k}_\bot^2\ge B_\rho^{-1}\approx 0.1\,\mbox{
GeV}^2\Rightarrow k_{\bot T}>300\mbox{ MeV}\,.\end{array}
\label{perplim}
 \ee

$\blacksquare$ In the {\bf numerical estimates} we use
the following parameters:

$\diamondsuit$ For the $\rho$ meson photoproduction we use
the HERA data, $\sigma_\rho\approx 12\,\mu$b (for the
diagonal in proton case, $p'=p$), $B_\rho\approx 10$
GeV$^{-2}$, $g_\rho^1\approx 1$,  $g_\rho^0\approx 0.1$.

$\diamondsuit$ For the odderon contribution we have no data.
The estimates given below and in Appendix show that at HERA
the odderon contribution would definitely dominate over the
other mechanisms if $\sigma_f\geq 1$ nb. Therefore, in order
to be able to make as strong conclusions as possible, we
take the value $\sigma_f=1$ nb.  (We hope that the real
value of $\sigma_f$ is significantly higher, e.g. 1 nb is
about 5\,\% from both the H1 experimental upper bound
\cite{H1} and the prediction of \cite{dosh}.) The slope
parameter $B_f$ for the $f_2$--meson photoproduction is also
unknown. For definiteness, we assume $B_f=B_\rho$.

$\Box$ The (charge symmetric) background is the sum of cross
sections obliged by Pomeron and odderon. Since the odderon
amplitude is considered to be very small, the background can
be approximated by the Pomeron $\rho$ contribution even far
from the $\rho$ peak, $d\sigma_{bkgd}/ dM^2\propto
|D_1(M^2)|^2$.

\begin{figure}[!htb]
   \centering
   \epsfig{file=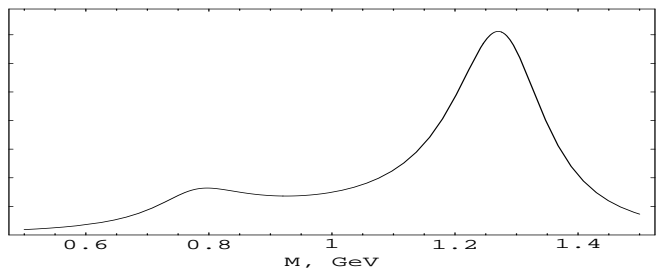,width=10cm,height=3.5cm}
      \caption{The local statistical significance
$SS_a(M^2)$ in arbitrary units.}
   \label{figSS}
\end{figure}

$\lozenge$ Let us consider values of statistical
significance (\ref{SB}) for cross sections averaged over
small interval of $M\pm\Delta M$, $SS(M^2)$. According to
eqs.~(\ref{i}) and (\ref{bckg}), for all asymmetries,
 \be
SS_a(M^2)\propto \fr{{\cal I}_{12}(M^2)}{|D_1(M^2)|}\equiv
\fr{Im( D^*_2D_1 e^{i\delta_{\Pom\od}})}{|D_1|} \le
|D_2|\,.\label{estss}
 \ee
Therefore the largest values of this  $SS(M^2)$ are located
near the $f_2$ peak. It is illustrated by
Fig.~\ref{figSS}, where local values of these $SS_a(M^2)$
are shown in arbitrary units. Hence,
to obtain the best value of $SS$, we consider signals and
background integrated over the reasonable mass interval
around the $f_2$ peak. A natural choice is
 \be
 M_f-\Gamma_f\;<M\;< M_f+\Gamma_f\,.\label{massint}
 \ee
Estimate (\ref{estss}) shows that the influence of
nonresonant background as well as tails of other resonances
in the Pomeron channel changes our estimates of $SS$ only
weakly. Since the overlap function exhibits no strong
dependence on phase difference $\delta_{\Pom\od}$ (see
Fig.~\ref{figov}), we use for estimates the value of the
integral over mass interval (\ref{massint}) at
$\delta_{\Pom\od}=0$:
 \be
\Delta{\cal I}=\int\limits_{M_f-\Gamma_f}^{M_f+\Gamma_f}
{\cal I}_{\rho f}(M^2)dM^2\approx 0.095\,.
 \label{overint}
\ee

Certainly, such estimate for interference with $S$--wave
$\pi^+\pi^-$ final states produced by odderon will show that
the corresponding signals are located near $f_0(600)$ and
$f_0(980)$ peaks, and these effects are negligible at $M>1100$
MeV.

$\Box$ The integration over the region (\ref{perplim}),
(\ref{massint}) results in background contribution
 \be
\sigma_B =\sigma_\rho\int
|D_1(M^2)|^2dM^2\times\left\{\begin{array}{c} 0.9\\
0.367\end{array}\right| \Rightarrow\left\{\begin{array}{lc}
428\mbox{ nb\  for } k_\bot>100\mbox{ MeV}\;& (FB)
\,,\\[2mm] 174\mbox{ nb\  for } k_\bot>300\mbox{ MeV}\;&(T)
\;\,.\end{array}\right.
     \label{bckg}\ee

$\Box$ We consider two cases of helicity structure of the
odderon amplitude. (Numerical factors in front of the
integrals below appear due to integration over the region
(\ref{perplim})).

$\triangledown$ {\it The $f_2$ meson is produced in the
state with helicity 1} (SCHC takes place also for $f_2$
production), $g_f^1\approx 1$. In this case the main
asymmetry will be FB (\ref{xiasym}), and the integration
over all variables in the region (\ref{perplim}),
(\ref{massint}) results in
 \be
\Delta\sigma_{FB} \approx 0.9\fr{3\sqrt{5}}{ 4}
\sqrt{\sigma_\rho \sigma_f}\cdot \Delta{\cal I} = 15.7\mbox{
nb}\Rightarrow SS\approx 7.5 \,.\label{FB}
 \ee

$\triangledown$ {\it The $f_2$ meson is produced via odderon
in the state with helicity 2}, $g_f^2\approx 1$ (SCHNC). In
this case the main asymmetry will be the transverse one
(\ref{vasym}). Performing integration in the region
(\ref{massint}), (\ref{perplim}), we obtain the same very
value as (\ref{FB})
 \be
\Delta\sigma_T = 0.507\cdot\fr{9\sqrt{5}}{16}
\sqrt{\sigma_\rho \sigma_f}\cdot \Delta{\cal I} = 6.6\mbox{
nb} \;\Rightarrow\;SS\approx 5\,.
 \ee

$\blacksquare$ These numbers are still very promising. This
offers certain confidence that the odderon signal is indeed
within the reach of the current experiments even with very
low value for the odderon--induced cross section and
relatively low luminosity (\ref{lum}).

\subsubsection{Possible discovery of the hard odderon
\cite{brod,hardpi}} \label{sechard}

$\blacksquare$ {\bm ${\cal T} =c\bar{c}$, $Q^2\approx 0$.}
The idea to discover odderon via the study of charge
asymmetry in the process $\gamma p\to c\bar{c}p$ was
proposed first in ref.~\cite{brod}. Since we deal here with
the production of heavy quarks, one can speak here about
{\it hard odderon}. In the specific calculation in
ref.~\cite{brod}, strong interaction of $c$ quarks in the
final state was neglected (in the spirit of the quark--hadron
duality). In this approximation, the overlap function
(\ref{i}) $ {\cal I}\Rightarrow \sin\delta_{\Pom\od}$
becomes $M$--independent, and the effect is strongly
underestimated near possible $c\bar{c}$ resonances.
Therefore the obtained in ref.~\cite{brod} estimates have
chances to be correct only far from the open charm
threshold, at $M\gg 2m_c$ (see sect.~\ref{secdual} for other
details).

$\blacksquare$ {\bf\bm ${\cal T}=\pi^+\pi^-$, $Q^2=2-3$
GeV$^2$}. The hard odderon can also be discovered via
observation of charge asymmetry (and related azimuthal
asymmetry) in the electroproduction of pion pairs by deeply
virtual photon \cite{hardpi}. The effect of longitudinal
virtual photons becomes essential in this region. When using
these results for real analysis, one should be careful.\\
 {\it i}) Their pQCD approach can be valid at $Q^2\gg
\Lambda^2_{QCD}$. Its validity for $Q^2\sim 2\div 3$ GeV$^2$
is unclear. For example, even for Pomeron at the considered
$Q^2$ the SCHC amplitude for transverse photon is not small.
These higher twist contributions can change results
strongly.\\
 ({\it ii}) The result contains sharp structure at
$|t|\approx 0.1$ GeV$^2$. It seems to be an artefact within
the pQCD approach, which can describe phenomena only with
averaging over interval of momenta wider than
$\Lambda_{QCD}^2$.\\
 ({\it iii}) The states of proton dissociation can be
different for Pomeron and odderon, which will reduce
interference.

$\bullet$ At a naive glance, "the main difference of studies
of the electroproduction process \cite{hardpi} with respect
to refs.~\cite{brod,GIN1} is to work in a perturbative
framework, which we believe enables us to derive more founded
predictions in an accessible kinematical domain"
\cite{hardpi}. Unfortunately, as we mention above and in
Appendix, in the considered kinematical region the
reliability of {\it numerical} calculations of papers
\cite{brod,hardpi} is low.

$\Box$ Besides, the discussed cross sections are very small,
so that the observation of these effects is hardly possible.
Indeed, for the $c\bar{c}$ photoproduction, even the Pomeron
mediated $c\bar{c}$ production cross section is small and
the efficiency of $c$--quark recording is low. For the
electroproduction of dipions by highly virtual photon the
observable cross section is also small since in this case
{\it(i)} the effective $\gamma p$ luminosity is much lower
than (\ref{lum}); {\it(ii)} the scale of cross section
itself is $\sim \alpha/Q^2$ instead of $\alpha/m_\pi^2$ for
soft case.

\subsection{\bm 20 MeV$\lesssim k_\bot\lesssim 100$ MeV,
$Q^2\approx 0$. Phases of the forward $\gamma p\to\rho p$
and $\gamma A\to \rho A$ amplitudes,
\cite{GINPom}}\label{secpom}

$\blacksquare$ The phase of the forward hadron elastic
amplitude ${\cal A}=|{\cal A}|e^{i\delta_F}$ at high energy
is an important object in hadron physics. In the naive
Regge-pole Pomeron model, this phase is given by the Pomeron
intercept as $\delta_F =(\pi/2)\alpha_\Pom$ (\ref{basform}).
The object, studied in modern experiments and named as
Pomeron, seems to be more complex. Measuring the phase of
this object appears useful in order to clarify its nature.

To the moment, phase of this type was measured in the only
type of experiments --- via study of Coulomb interference in
$pp$ or $\bar{p}p$ elastic scattering. These experiments
require measurement of extremely low scattering angles,
which becomes practically impossible at high enough
energies.

$\blacksquare$ The study of charge asymmetry in the process
$\gamma p\to \pi^+\pi^- p$ (or $ep\to e\pi^+\pi^- p$) at
very low transverse momenta of the produced dipion provides
new method to measure this type of phase \cite{GINPom}.
For this purpose, we suggest to study  the charge asymmetry
in the same mass interval (\ref{massint}) and at very small
$k_\bot$,
 \be
k_{min}= 20\mbox{ MeV}< k_\bot<k_{max}\approx 100 \mbox{
MeV}\, \Rightarrow |t|_{cut}=0.01\mbox{ GeV}^2\,.
\label{perppom}
 \ee
We speak here about the total transverse momentum of dipion
with effective mass $>1$ GeV. In this case typical
transverse momenta of separate pions are $\sim 500$ MeV,
i.e. they are measurable well. In the eq.~(\ref{perppom})
lower limit $k_{min}$ is determined by accuracy in the
measurement of $p_{\pm\bot}$. The upper limit $k_{max}$
describes the region where the Coulomb (Primakoff)
contribution is essential while alternative hadronic
contributions are inessential. At $k_\bot<k_{max}$ the
contributions of the proton excitations are negligible for
both discussed mechanisms. Besides, in this region the
form-factor effects are inessential, and one can treat
proton as a point-like particle.

$\bullet$ In the region (\ref{perppom}) the C--even dipion
is produced mainly via photon exchange (Primakoff effect),
studied well at \epe\ collisions. It is described with very
high precision by the equivalent photon approximation (see
for details \cite{BGMS}). Similar to eqs.~(\ref{ampl}), one
can write this amplitude in the form
 \be
{\cal A}_\gamma= Cg_\gamma^{\lambda_f}\fr{|{\bf
k}_\bot|}{\vec{\bf k}_\bot^2+Q^2_m} D_R(M^2){\cal
E}^{J,\lambda_R}_{\lambda_{\gamma}}\;\mbox{ with }
Q^2_m\approx\left(\fr{m_pM^2}{s}\right)^2\label{Primampl}
 \ee
with normalization factor $C$ given by the QED result
involving the two photon width $\Gamma_{\gamma\gamma}$
 \be
d\sigma_f=\fr{8\pi\alpha
\Gamma_{\gamma\gamma}(2J+1)}{M^3}\cdot \fr{\vec{\bf
k}_\bot^2 d\vec{\bf k}_\bot^2} {(\vec{\bf
k}_\bot^2+Q^2_m)^2}\,. \label{Prim}
 \ee
The total cross section of the $f_2$ meson production and
the cross sections in different kinematical regions for HERA
case are
 \bear{c}
 \sigma_f^{tot}\approx \fr{8\pi\alpha
\Gamma_{\gamma\gamma}(2J+1)}{M^3}
\left(\ln\fr{m_\rho^2}{Q_m^2}-1\right)= 8\mbox{
nb}\,,\\[4mm]
 \sigma_f(k_\bot\le 100{ MeV})>7\mbox{
nb}\,,\qquad\sigma_f(k_\bot\ge 300\mbox{ MeV})\approx
0.2\mbox{ nb}\,.\end{array}
 \label{Primnumb} \ee
Large cross section is concentrated in the narrow region
near the forward direction. That is the ground for our
choice of the region (\ref{perppom}) for the study of the
discussed effect.

For the considered collision of almost real photons only two
values of dipion helicity are allowed by the conservation
laws, 0 and 2, i.e. $g_\gamma^{\lambda=1}=0$. For the $f_2$
meson ($J=2$) data give $g_\gamma^{\lambda=2} \approx 1\gg
|g_\gamma^{\lambda=0}|$.

$\bullet$ The C-odd dipions are produced via diffractive
Pomeron mechanism through the $\rho$--meson like state,
mainly in the state with helicity 1 (SCHC), as it was
discussed above. This amplitude is described by
eqs.~(\ref{ampl}). At $k_\bot <100$ MeV we have
$B_\rho|t|<0.1$. Therefore, one can neglect $t$ dependence
of strong interaction ("Pomeron") amplitude.

The background contribution is the sum of Pomeron and
Primakoff effects,
 \be
 \sigma_B\equiv \sigma_\rho+\sigma_f=\left[470B_\rho (k_{max}^2-
 k_{min}^2)+
 0.45\ln\fr{k_{max}^2}{k_{min}^2}
\right] \mbox{nb}\approx 47+1.5=48.5\mbox{ nb}\,.
\label{bckgPr}
 \ee

$\blacksquare$ The charge asymmetry is calculated now just
as for the odderon case. Certainly, in this case the overlap
function ${\cal I}_\Pom$ is determined by eq.~(\ref{i}) with
natural change of $\delta_{\Pom O}$ to quantity $\delta_F$.
In accordance with discussion in sect.~\ref{secodd}, the
main asymmetry is transverse one, and similarly to
(\ref{vasym}), but in contrast with the odderon case,
asymmetry increases with decreasing of $k_\bot$, and (we set
$g_\rho^1g_f^2\approx 1$)
 \be
\fr{d\sigma_T^{Pr}}{dM^{2}\, d\bf{k}_\bot^2\,dv}
=\fr{3\sqrt{5}}{8}C {\cal I}_{\Pom}(M^2)
\fr{\sqrt{\sigma_\rho B_\rho}}{{|k_\bot|}} v(1+v^2)
\,.\label{Pomasym}
 \ee

$\Box$ To estimate the opportunity to observe the asymmetry
discussed, it is useful to calculate the overall effect,
i.e. asymmetry, integrated over $M^2$, $k_\bot$ and $v$ in
the regions (\ref{perppom}), (\ref{massint}). In this
estimate for the integral of overlap function one can use
quantity (\ref{overint})
 \be
 \Delta\sigma_T=
\fr{9\sqrt{5}}{8} C\sqrt{\sigma_\rho
B_\rho}(k_{max}-k_{min})\cdot\Delta{\cal I}\approx 5.6\mbox{
nb} \,.\label{PomPrnumb}
 \ee

For the $\gamma p$ effective luminosity (\ref{lum}) it
results in statistical significance $SS\approx 7.5$. This SS
value is sufficient for the observation of an effect.
However, to extract the phase under interest, one should
study in detail the mass dependence of asymmetry. For this
purpose it is useful to have larger effective $\gamma p$
luminosity integral, and this luminosity, necessary for
reasonable precision, should be estimated at the stage of
planning the experiments.

$\diamondsuit$ The simplified model for $\rho$ and $f_2$
amplitudes, used for estimates, seems to be too rough. The
contributions of other resonances and nonresonant background
should be included. Instead of such calculation (which has
many sources of ambiguity), the precise phase of the C--even
dipion production amplitude can be found via the study of
charge asymmetry in the $e^+e^-\to e^+e^-\pi^+\pi^-$ process
(sect.~\ref{eepipip}) in the same effective mass region.

\subsubsection{\bm $eA\to e\pi^+\pi^-A$. The nuclear "Pomeron" phase}

The study of charge asymmetry of pions in the $eA\to
e\pi^+\pi^-A$ collisions with heavy nuclei (at the future
ERHIC and THERA colliders) provides an opportunity to
measure the Pomeron phase $\delta^A_F$ for a nuclear target.
(Generally $\delta^A_F\neq \delta_F$ for proton target).

$\blacksquare$ When the ultrarelativistic heavy nuclei
collide, there is the region of momenta transferred from one
of them, which is so small that it does not destroy this
nucleus --- {\it ultra-peripheral collisions} (UPC). In this
region the electromagnetic interaction of the nucleus is
coherent, its strength is defined not by the fine structure
constant $\alpha\ll 1$, but by the quantity $Z\alpha\sim 1$.
Therefore, the electromagnetic interactions become
comparable with strong interactions (or even become stronger
due to Coulomb pole). These interactions are characterized
by the nuclear charge $Ze$ and formfactor $F_A(q^2)$ with
scale of the $q^2$ dependence $\Lambda^2\sim 1/R_A^2$. For
heavy nuclei considered here $\Lambda\approx 60$ MeV.

For the coherence (UPC), one should have $|q^2|<\Lambda^2$.
The transferred momentum $q$ is related to its transverse
component $q_\bot$ as
\be
   -q^2=q^2_m+q_\bot^2\,, \quad
  q^2_m=\omega^2/\gamma^2_A\,,\label{transfmom}
  \ee
where $\gamma_A$ is the nuclear Lorenz factor and $\omega$
is the energy transferred from the nucleus. Therefore, the
transverse momentum end energy transferred to the dipion are
limited as
 \bes\label{2UPCnumb}\bea
 &k_\bot \lesssim \Lambda\approx 60\mbox{ MeV}\,,&\label{2UPCtr}\\
&\omega<\Lambda\gamma_A
 =\left\{\begin{array} {l}
 6\;\;\mbox{ GeV for ERHIC e-Au},\\
180\mbox{ GeV for LHC
e-Pb}.\end{array}\right.&\label{2UPClong}
 \eea\ees

For the virtual photon $q_e$, emitted by electron we require
only the limitation (\ref{2UPCtr}) for $q_{e\bot}$. It
allows to transmit the limitation (\ref{2UPCtr}) to the
produced dipion. For this photon we have $-q_e^2=
[Q_m^2+q_{e\bot}^2]/(1-\omega/E)$ with $Q_m\approx
m_eM^2/s$. Therefore, this photon can be treated as
quasireal in respect to the production of hadron system
$\cal T$ (in this case $|q_e^2|<\Lambda_{\cal T}^2\sim
m_\rho^2$), and we treat the UPC $eA\to e{\cal T}A$ as {\bm
the $\gamma A\to {\cal T} A$} processes.

$\blacksquare$  We suggest to look for dipions at the mass
interval (\ref{massint}) and in the interval of $k_\bot$
(instead of (\ref{perppom}))
 \be
 20\mbox{ MeV}=k_{min}< k_\bot<k_{max}\approx 60 \mbox{ MeV}\,
\Rightarrow |t|_{cut}=0.0036\mbox{ GeV}^2\,.
\label{perppomA}
 \ee
Since the mean number of nuclear collisions per bunch
crossing in ERHIC will be less than 1, these UPC can be
isolated in events with this kinematical limitation and {\bf
without other particles in the detector}.

Subsequent estimates are similar to those in the
previous subsection.

The C--even dipions are produced via photon exchange
(Primakoff effect). This production is described by
eqs.~(\ref{Primampl}) with additional factor
$ZF_A(k_\bot^2+Q^2_m)$. For estimates, we write the
form-factor in the form $F_A(Q^2)=1/(1+Q^2/\Lambda^2)$.

At large enough $\gamma A$ energies\fn{ At ERHIC --- with
large enough longitudinal momentum of the dipion directed
along the initial electron motion or at THERA.} C-odd
dipions are produced via diffractive $\rho$--meson--like
production, mainly in the state with helicity 1. The
approximation for the Pomeron amplitude (\ref{ampl}) is
valid with additional factor $A^{2/3}$ and the change
$\sigma_\rho B_\rho\to \sigma_\rho^A B_\rho^A$. The values
$\sigma_\rho^A$ and $1/B_\rho^A$ are smaller than the
corresponding quantities for proton \cite{Ryskin}. In the
numerical estimates we write $\sigma_\rho^A
B_\rho^A=k^2\sigma_\rho B_\rho$ with coefficient $k\sim 1$.
In any case, in the region (\ref{perppomA}) the $t$
dependence of "Pomeron" amplitude is negligible.

For the collision of electron with $E=100$ GeV and $Au$
nuclei with $\gamma_A=109$ we obtain the total cross section
of the $f_2$ production and the cross sections in the region
(\ref{perppomA})
 \bear{c}
 \sigma_f^{tot}\approx Z^2\fr{8\pi\alpha
\Gamma_{\gamma\gamma}(2J+1)}{M^3}
\left(\ln\fr{\Lambda^2}{Q_m^2}-1\right)\approx 1.5Z^2\mbox{
nb}\,,\\[4mm]
 \sigma_f(k_\bot\ge 20{ MeV})\approx 0.6Z^2\mbox{
nb}\,.\end{array}
 \label{PrimnumbA} \ee

The background contribution is the sum of the Pomeron and
Primakoff effects,
 \be
 \sigma_B=\left[470 A^{4/3}k^2
 B_\rho (k_{max}^2-k^2_{min})+
 0.3 Z^2\ln\fr{k_{max}^2}{k_{min}^2}
\right] \mbox{nb}\approx
\left(16k^2A^{4/3}+0.66Z^2\right)\mbox{ nb}\,.
\label{bckgPrA}
 \ee

{\bf The charge asymmetry} is calculated just as for the
$\gamma p$ case with similar overlap function ${\cal
I}_\Pom$. In accordance with discussion in
sect.~\ref{secodd}, the main asymmetry is transverse and,
similarly to eq.~(\ref{Pomasym}), the asymmetry increases
with decreasing $k_\bot$:
 \be
\fr{d\sigma_T^{Pr}}{dM^{2}\, d\bf{k}_\bot^2\,dv}
=ZA^{2/3}k\fr{3\sqrt{5}}{8}C {\cal I}_{\Pom}(M^2)
\fr{\sqrt{\sigma_\rho B_\rho}}{{|k_\bot|}} v(1+v^2) \,.
 \ee

$\Box$ The overall asymmetry is an integral over $M^2$,
$k_\bot$ and $v$ in the regions of eqs.~(\ref{perppomA}),
(\ref{massint}). With the integral of the overlap function
given by the quantity (\ref{overint}), we have
 \be
 \Delta\sigma_T=
ZA^{2/3}k\fr{9\sqrt{5}}{8} C\sqrt{\sigma_\rho
B_\rho}(k_{max}-k_{min})\cdot\Delta{\cal I}\approx
2.8ZA^{2/3}k\mbox{ nb} \,.\label{PomPrnumbA}
 \ee
Since $A^{2/3}\approx Z$, the S/B ratio is almost the same
for different nuclei. However, the value of SS (\ref{SB})
increases $\propto Z$ with $Z$ growth at fixed luminosity
integral $\cal L$. Since the Pomeron contribution dominates
in the background, the factor $kA^{2/3}$ disappears from the
estimate of SS. For $Au$ nuclei even at low ${\cal L}\approx
1$ nb$^{-1}$ we have a very good $SS$ value $\approx 14$.

\subsection{Study of double Pomeron exchange in UPC}

For pA or AA collisions of ultrarelativistic heavy
nuclei A, UPC provides a good tool to study the double
Pomeron phenomena. Under the kinematical conditions
eq.~(\ref{2UPCnumb}) for the LHC and (\ref{2UPCtr}) for the
$pA$ RHIC and (with bad precision) for HERA-B, the main
mechanisms for the dipion  production in the process
$A_1A_2\to A_1A_2\pi^+\pi^-$ are
 \bear{l}
\bullet\mbox{ Pomeron -- Pomeron }\;(A_1\Pom A_1\,,\;\;
A_2\Pom
 A_2)\;\otimes\; \Pom\Pom\to \pi^+\pi^-\,,\\
\bullet\mbox{ Photon -- photon } (A_1\gamma A_1\,,\;\;
A_2\gamma
 A_2)\;\otimes\; \ggam\to \pi^+\pi^-\,,\\
\bullet\mbox{ Pomeron -- photon }(A_1\Pom A_1\,,\; A_2\gamma
A_2)\;\otimes\; \Pom\gamma\to \pi^+\pi^-\,.
 \end{array}
 \label{mechanismsAA}\ee

Similar to the estimates in the previous section, one can
expect the Pomeron-Pomeron contribution to dominate, and
the main charge asymmetry to appear due to interference of
Pomeron--Pomeron and Pomeron--photon amplitudes. After the
studies of the Pomeron phase in $eA$ collisions, these
investigations in nuclear collisions open the door to a
detailed study of the Pomeron-Pomeron amplitude, which is
now poorly understood.

\subsection{\bm $Q^2\gtrsim 1000$ GeV$^2$, $k_\bot\lesssim 1$ GeV.
Study of coupling of the axial current to the Pomeron \cite{GIV}}

At large electron scattering angles ($p_{\bot e}\gtrsim 30$
GeV), the interaction of electron with proton via $Z$--boson
exchange (mainly {\it axial current}) becomes essential in
addition to the standard photon exchange {\it (vector
current)}. In this region we suggest to consider also dipion
final state with large rapidity gap and with no specific
final state for proton excitation. In this case the
interaction of vector and axial currents with proton (via
the Pomeron exchange) produces C--odd and C--even final
states respectively. Both amplitudes are described by
approximation of form (\ref{ampl}). The content of final
state $p'$ is identical in both cases providing as complete
interference as possible.

Before a detailed analysis of data it is difficult to say
whether any resonant states or nonresonant background
dominate in this region. For preliminary estimates at 1.1
GeV$<M<1.5$ GeV one can use calculations of $\rho$ and $f_2$
production in pQCD (with massless quarks) in 2-gluon
approximation. For $\rho$ production these amplitudes are
written in \cite{GIvQ}. Calculation of the $f_2$
production by axial current can be made in this very manner
by using the approach of \cite{GIvod} as well. One can
expect that this approximation gives correct shape of the
charge asymmetry while the value of effect (and background)
will be enhanced due to the Pomeron enhancement in
comparison with two-gluon approximation.

The interference of these amplitudes results in charge
asymmetry effect $\sim (Q^2/M_Z^2)$ with overlap factor
which is different from that in (\ref{i})
 \be
  {\cal I}^P_{\rho f}(M^2) =
Re\left[D_\rho D_f^\dagger \right]= Re\left(\fr{\sqrt{m_\rho
m_f\Gamma_\rho\Gamma_f
Br(f_2\to\pi^+\pi^-)Br(\rho\to\pi^+\pi^-)}}
{\pi(M^2-m^2_\rho+ i m_\rho\Gamma_\rho) (M^2-m_f^2-i m_f
\Gamma_f)}\right)\,.
 \label{iP}
 \ee
\begin{figure}[htb]
\label{figover2}
\begin{center}
  \includegraphics[bb=155 8 441 186,width=0.5\textwidth,height=0.17\textheight,clip=true]
  {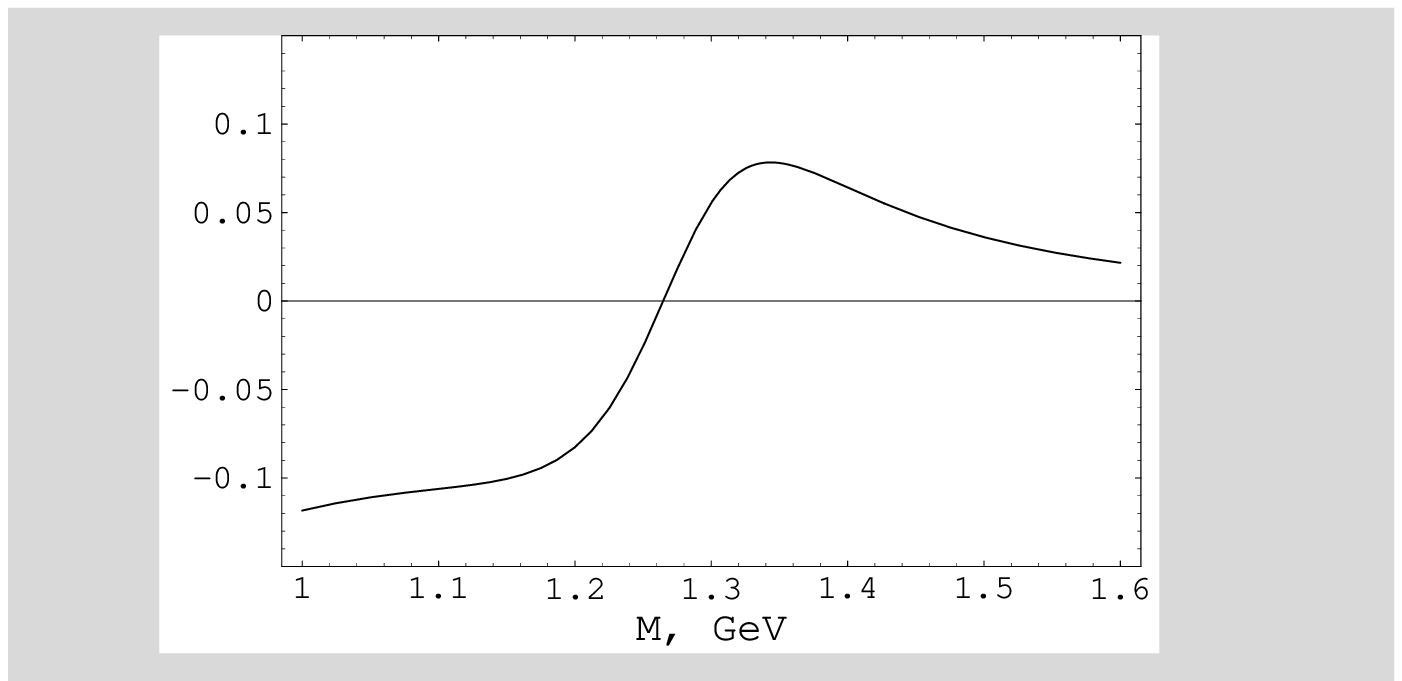}\\
  \caption{\em The $\rho-f_2$ overlap function
           for $ep\to e\pi^+\pi^-X$  at large $Q^2$.}
  \end{center}
\end{figure}
In particular, near the $f_2$ pole the contribution of $f_2$
is almost imaginary while the contribution of $\rho$ is
roughly real. Therefore, this overlap function changes its
sign at $M\approx M_f$. The peaks in the mass distribution
are disposed at the distance $\sim \Gamma_f$ from this pole.
The idea of estimate of Statistical Significance
(\ref{estss}) is valid in this case as well. Finally for the
reasonable and careful estimate of SS one should perform
mass integration over the region (instead of
eq.~(\ref{massint}))
 \be
 {\cal D}_R:\;\; M_{f_2}+\vak\Gamma_{f_2}<M
 <M_{f_2}+(1+\vak)\Gamma_{f_2}\,,
\qquad (\vak\sim 0.2\div 0.5)\, . \label{massintP}
 \ee
The signal below 1.2 GeV also includes contributions from
$f_0$ resonances.

\section{Breaking of quark--hadron duality}\label{secdual}

It is usually assumed that for heavy quarks the
quark--hadron duality (Q-HD) works well (at least in
average). However, this Q-HD is violated strongly in the
charge asymmetry phenomena due to the final state
interaction --- FSI.

$\bullet$ Let us remind that the charge asymmetry of muons
in the process $\epe\to \epe\mu^+\mu^-$ (Fig.~2, right
panel) differs strongly from that of pions (Fig.~2, left
panel). For muons $|\Delta\sigma^\mu_T|\gg
|\Delta\sigma^\mu_{FB}|$ while for pions (QED)
$|\Delta\sigma^\pi_{FB}|\gg |\Delta\sigma^\pi_T|$. At the
first glance, the charge asymmetry for the (point-like)
heavy quarks should be mainly transverse as that for muons.
In reality, hadronization transforms quarks into mesons with
spin 0 and 1, and one can expect that the FB asymmetry will
be the largest near the threshold like that for pions. This
change of type of the charge asymmetry will be a clear
signal of the Q-HD breaking.

$\bullet$ The main source of observed $D$--mesons is the
decay of $c\bar{c}$ resonances. These are $J=1$ states of
$c\bar{c}$ system, produced via bremsstrahlung production
for \epe collision or via Pomeron exchange for $\gamma p$
case. In the same region the $J=0$ and $J=2$ resonance
states of this $c\bar{c}$ system should be produced by two
photons or via odderon exchange respectively. All these
resonances are not very narrow. Overlapping of these
resonances should give essential contribution to the
charge asymmetry as it was discussed for pions.

For the $\gamma p$ collision for point-like quarks the
charge asymmetry would be suppressed by a small factor
${\cal I}= \sin\delta_{\Pom\od}$ obtained in
ref.~\cite{brod}. In reality, due to FSI -- resonance
production, we expect picture which is similar to pion pair
production where this small factor is eliminated due to
additional phase shift given by product of two Breit--Wigner
factors related to different resonances (see eq.~(\ref{i})
and Fig.~\ref{figov}) .

\section{ Weighted  structure functions}

Now we consider the {\bf deep inelastic scattering} of
electron with momentum $p_e$ on the proton with momentum
$p_p$ (DIS). Let $w$ be some charge asymmetric quantity
determined for all observed particles, $\hat{C}w=-w$.

For example, {\sl let $x_j$ be standard light cone variable
for each produced particle $j$, i.e.  $ x_j=
(E_j-p_{jz})/E_e\equiv 2p_jp_p/s$. We define the weight
factor $w_\xi= \sum\limits_{j=+}x_j- \sum\limits_{i=-}x_i
\equiv 2(p_+-p_-)p_p/s$ (cf. (\ref{defbasxi}))}. (The first
term here is the sum over all positively charged
secondaries, the second one is the sum over all negatively
charged secondaries.)

$\blacksquare$ The suggested {\it weighted structure
functions --- WSF --- are described via data in the same
manner as the usual structure functions for DIS but with
weight factor $w$ like $w_\xi$  {\bf for each event}}.
(Certainly, the standard polarization analysis of an initial
state can be added to this definition.) These weight factors
bring the charge asymmetry into standard definitions. In the
standard language they are
 \be
 W_{\mu\nu}^{C,i}(p,q)=2\pi^2\sum\limits_X\int d^4 z
<p|J_\mu(z)|\hat{W}^C|X><X|J_\nu(0)]|p>e^{iqz} \label{strf}
 \ee
with charge odd operators $\hat{W}^C|X>=w|X>$. It is
desirable to have such form of weight $w$, to minimize
influence of the target specifics. {\sl(This feature works
in our example $w_\xi$ since the quantities $x_j$ are small
for secondaries $j$, flying along initial proton.)}

$\diamondsuit$ If the operator $\hat{W}^C$ acts for hadron
and quarks similarly and is measurable explicitly in each
event, these WSF will give us new information about
quark--gluon structure of matter at small distances.

$\diamondsuit$ If the operator $\hat{W}^C$ acts for hadron
and quarks in different ways, the WSF proposed will be
sensitive to the details of confinement as well.

$\blacksquare$ {\bf Useful points for the WSF.}

$\bullet$ It is well known that the (multi)gluon exchange
effects cannot be seen in the standard $W_3$-like functions.
Indeed, for the (multi)gluon colorless exchange with proton,
the function (\ref{strf}) corresponds to the interference of
C-even (Pomeron-like) exchange and C-odd (odderon-like)
exchange, which produces in the collision with photon the
final states of the opposite C-parity. Therefore, this
contribution disappears in the standard structure functions
(see \cite{kwecz} for lower approximation). Respectively,
this contribution remains in the weighted object.

$\bullet$ Usually, the contributions of the vector current
$J_V$ (mainly from photon exchange) and axial current $J_Z$
(from $Z$ exchange) in the structure functions are summed
without interference, and the second contribution is small
fraction ($\sim (Q^2/M_Z^2)^2$) at $Q^2< (70\;GeV)^2$. The
difference of cross sections for the left-hand and
right-hand polarized electrons (like $W_3$ structure
function) is
 \be
d\sigma^L-d\sigma^R\propto Re(J^*_VJ_A)\,.
 \ee
That is C--odd quantity. Therefore, using some C-odd
weight function in WSF makes this interference clean.
According to our experience in Higgs physics, this WSF
should become large enough at $Q^2>1000$ GeV$^2$
($p_{e\bot}>30$ GeV).

\section{\bm  $e\gamma\to eWW$. Strong interaction in the Higgs sector}

The possible  strong interaction in the Higgs sector can be
seen as that of longitudinal $W$'s, in particular, in the
process $\ggam\to W_LW_L$. The experience with $\ggam\to
\pi^+\pi^-$ makes the following picture very probable:
strong interaction modifies weakly the cross section near
the threshold in comparison with its QED value but the phase
of amplitude reproduces that of strong interacting $W_LW_L$
scattering and can be not small. This makes strong
interaction in the Higgs sector badly observable in the
cross sections below  expected masses of $WW$ resonances,
about 1.5--2 TeV.

The charge asymmetry in the process $\egam\to eWW$ is given
by interference of two--photon and one-photon production (as
in $\epe\to\epe \pi^+\pi^-$) and by interference of photon
and $Z$ boson exchanges. The different interferences
dominate in different regions of the final phase space (in
dependence on the transverse momentum of the scattered
electron and its energy).

The study of charge asymmetry in the process considered can
be the key to the discovery of this strong interaction at the
relatively low energy of TESLA (0.8 TeV) much below possible
resonance production, since it is sensitive to $W_LW_L$
scattering phase shifts \cite{GSD,AGP}. To distinguish
between this charge asymmetry and that for the lepton
final states, discussed in the next section, it is
necessary to consider quark decays of W-bosons.

\section{\bm Polarization charge asymmetry in \ggam\
collisions}

Let us consider {\it the charge asymmetry  in \ggam\
collisions, coming from the definite polarization of the
initial state} in the processes like $\ggam\to WW\to {\cal
T}\nu\bar{\nu}$, $\ggam\to \tau^+\tau^-\to {\cal
T}\nu\bar{\nu}\nu\bar{\nu}$ or $\ggam\to \chi^+\chi^-\to
{\cal T}\nu\bar{\nu}\chi_0\bar{\chi}_0$ with ${\cal
T}=\mu^+\mu^-$ (or ${\cal T}=\mu^+e^-$, $\mu^-e^+$, $\epe$).
(Here $\chi^\pm$ is chargino and $\chi_0$ is neutralino --
LSP.) These processes will be studied at the Photon
Colliders \cite{GKST,TESLATDRVI} where high energy photons
will be prepared mainly in the states with definite helicity
$\lambda_i\approx \pm 1$. For discussion, we distinguish the
initial states with $\lambda_1\,,\;\lambda_2=\pm 1$.

The QED cross sections of pair production $\ggam\to WW$,
$\ggam\to \tau^+\tau^-$, etc. depend on the product
$\lambda_1\lambda_2$ only and exhibit no charge asymmetry
(due to P-invariance of electromagnetic interactions).
However, the helicity states of the intermediate $W^\pm$,
$\tau^\pm$ or $\chi^\pm$ depend separately on photon
polarizations. In the subsequent decay of these $W$, $\tau$
or $\chi^\pm$, the P-parity is not conserved. This gives
correlations between spin of intermediate $W$, $\tau$ or
$\chi^\pm$, etc. and the momentum of the single particle
observable in the final state of this decay (in our example
--- the muon). Due to opposite directions of polarizations
of intermediate particles and antiparticles, final
distributions of observed $\mu^+$ and $\mu^-$, etc. become
different, and the charge asymmetry arises. Note that this
asymmetry is absent for massless intermediate particles due
to helicity conservation. Therefore, the value of effect
increases with the mass of intermediate particle ($\tau$ or
$W$ or $\chi^\pm$ in our examples).  Certainly, the
observable effect summarizes effects from various
intermediate states. So that the detailed study of charge
asymmetry, related to different mechanisms, in different
regions of final phase space is necessary.

$\bullet$ The initial states $(\lambda_1,\lambda_2)=+-$ and
$-+$ choose preferable direction of the collision axis.
Therefore, in this case we expect FB asymmetry of final
muons. It has opposite signs for $+-$ and $-+$ initial states.

$\bullet$  The initial states $(\lambda_1,\lambda_2)=--$ and
$++$ choose preferable direction of rotation (left or right,
respectively) but the opposite directions of the collision
axis are equivalent. Therefore, we expect here T asymmetry
of final muons, but not FB asymmetry. It has opposite signs
for $++$ and $--$ initial states. Certainly, this FB
asymmetry will be smoothed due to nonmonochromaticity of
incident photons.

\begin{figure}[htb]
\begin{center}
{\bf\bm
\includegraphics[width=0.24\textheight,height=0.49\textwidth,angle=-90] 
{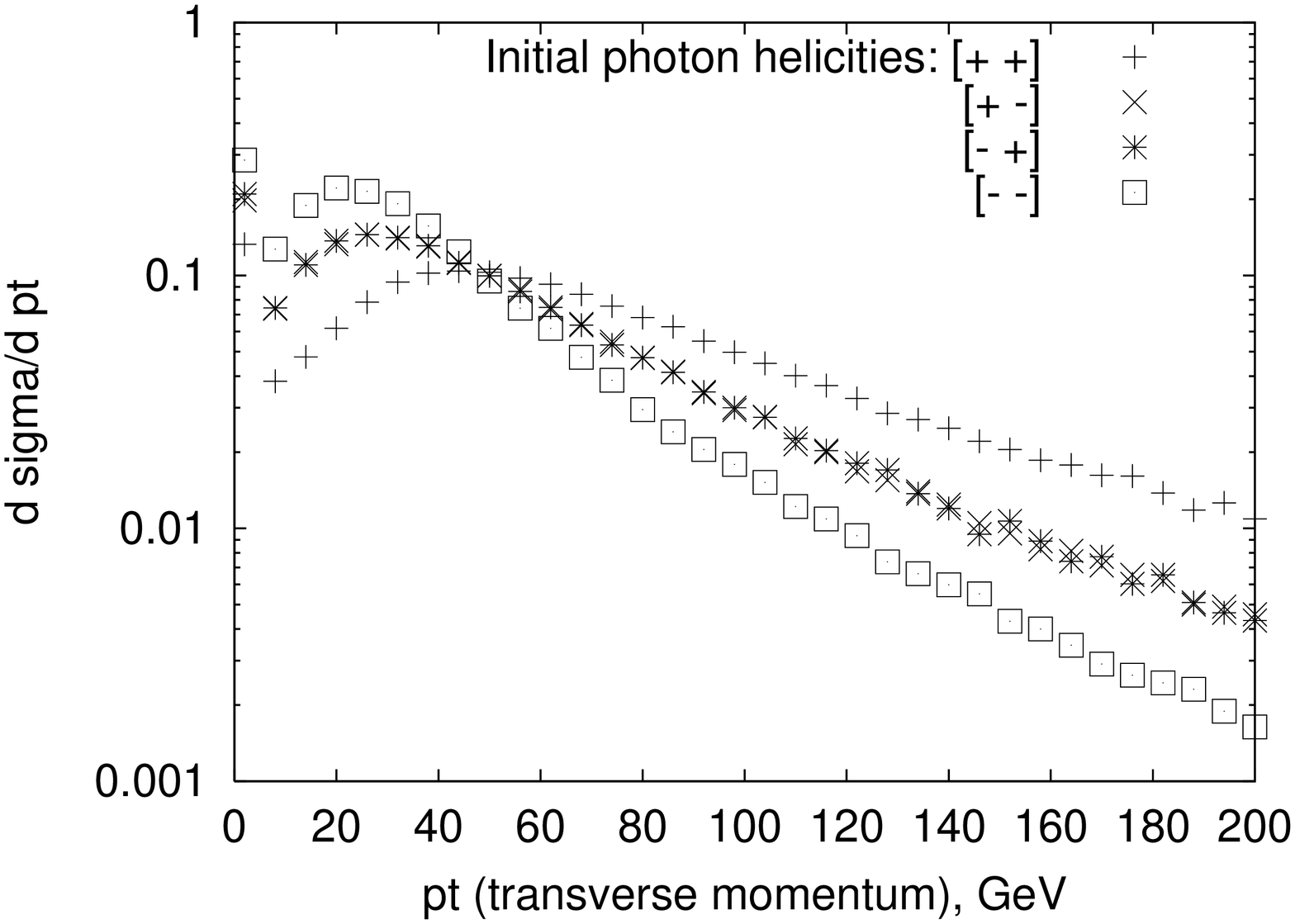}
\includegraphics[width=0.24\textheight,height=0.49\textwidth,angle=-90]
{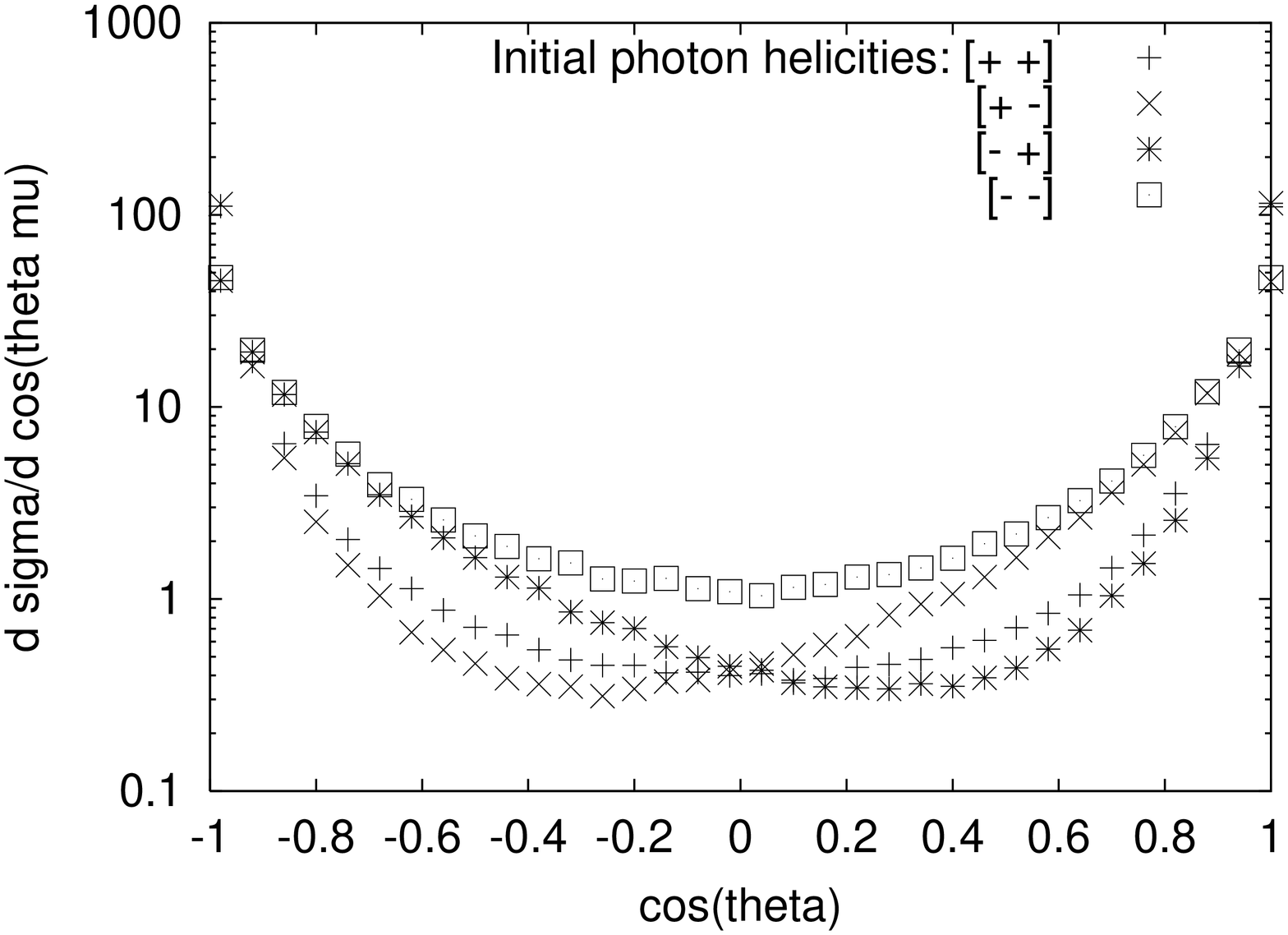}
}
 \caption{\em Cross-sections ${d\sigma}/{dp_{\mu\bot}}$
 (left) and ${d\sigma}/{dcos(\theta_{\mu})}$ (right)
 of the process $\gamma\gamma\to W\mu\bar{\nu}$}
\end{center}
\label{fig:8}
\end{figure}

These features are clearly seen in Fig.~8 where muon
distributions in the process $\ggam\to W^+\mu^-\bar{\nu}$
are shown for different initial photon states. One can see
here that the transition from +- to -+ initial states
changes transverse distribution of muons, that corresponds
to the T asymmetry for $\ggam\to \mu^+\mu^-\nu\bar{\nu}$
process. The transition from ++ to -- initial states does
not change transverse distribution of muons but changes its
angular distribution (due to change of distribution in
longitudinal momentum), that corresponds to FB asymmetry in
$\ggam\to\mu^+\mu^-\nu\bar{\nu}$ process.

These asymmetries and the quality of resonance approximation
for their description were  analyzed in detail in
ref.~\cite{polasym} for the simplest EW process $\ggam\to
WW\to \mu^+\mu^-\nu\bar{\nu}$ within SM together with
asymmetries in the process $\ggam\to\tau^+\tau^-\to
\mu^+\mu^-\nu\bar{\nu}\nu\bar{\nu}$ considered in the
resonant approximation. The similar FB asymmetry was
considered in Ref.~\cite{chargino} for the $\ggam\to
\chi^+\chi^-\to\epe \nu\bar{\nu} \chi_0\bar{\chi}_0$
process\fn{ Note that the discussion of \cite{chargino} of
variations related to change of helicities of initial laser
photon and electron at $e\to\gamma$ conversion is completely
related to variation of initial \ggam\  state discussed in
detail in \cite{GKST}. Effects discussed there can be
obtained by variation of initial beam energy.}.

The study of this charge asymmetry can be a good tool for
investigation of different effects of the New Physics (anomalous
triple and quartic interactions of gauge bosons, strong
interaction in Higgs sector, SUSY, ...). In this task the
mentioned increase of charge asymmetry with the growth of
mass of an intermediate particle will be useful.

Unfortunately, the mentioned discussions of $\ggam\to
\tau^+\tau^-\to \mu^+\mu^-\nu\bar{\nu}\nu\bar{\nu}$ and
$\ggam\to \chi^+\chi^-\to \epe\nu\bar{\nu}
\chi_0\bar{\chi}_0$ are very preliminary since they use
resonance approximation for description of final states with
6 particles. This approach neglects a huge number of other
diagrams with the same final state
  (for the process $\ggam\to
\mu^+\mu^-\nu\bar{\nu}\nu\bar{\nu}$ the resonant
approximation deals with 2 diagrams instead of 184 in SM).
 In the complete form this problem is difficult for
computing. (Of course, many omitted diagrams do not
contribute to the asymmetry.) The development of
corresponding computing algorithms is necessary.

Note that in these problems the system ${\cal
T}(=\mu^+\mu^-$ in our example) is organized from two
particles of different origin. Therefore, the useful
variables for description of these asymmetries become
not dimensionless (\ref{asymdef}), but their
dimensional analogs, for example, for the transverse
asymmetry $\tilde{v}= \mb{p}_{+\bot}^2-\mb{p}_{-\bot}^2-K_-
{\mb k}_\bot^2$.

\section*{Appendix. The  status of the odderon vs.
Pomeron\fn{ In this Appendix I follow discussion of
ref.~\cite{GIN1}.}}

The Pomeron and odderon are treated as the reggeons, that
have vacuum quantum numbers with the only difference: while
Pomeron is C--even, the odderon is C-odd, similarly to the
photon. Pomeron exchange describes small angle elastic and
total cross sections at high energies. Odderon is
responsible, e.g., for the difference $\sigma_{pp}^{tot}
-\sigma_{p\bar{p}}^{tot}$ at high energies
\cite{Gribov,Luk}.  Assuming the Pomeron and odderon are
Regge poles, their contributions to the scattering amplitude
$AB\to CD\;$\ must have the form (\ref{basform}),
(\ref{bashel}). Because  $|\sigma^+ - \sigma^-|
<\sigma^++\sigma^-$, one should be $\alpha_\od \le
\alpha_\Pom$. Since the intercepts $\alpha_\Pom$ and
$\alpha_\od$ are close to 1, the Pomeron exchange amplitude
is predominantly imaginary, while the odderon exchange
amplitude is predominantly real.

Within perturbative QCD, the Pomeron and odderon are based
on two--gluon and $d$--coupled three--gluon exchanges in
$t$--channel respectively \cite{Bart}. Hence, both the
Pomeron and odderon intercepts are close to the gluon spin,
$\alpha_\od,\,\alpha_\Pom \sim 1$. The experimental data and
BFKL calculations show that the Pomeron intercept
$\alpha_\Pom(0)>1$. The theoretical estimations for odderon
intercept vary with the date of preparation of paper
($\alpha_{\od}(0)=0.94\to 0.96\to 1\to ?$).

The cross section difference $\sigma_{pp}-
\sigma_{p\bar{p}}$ is less than the experimental
uncertainties. The diffractive photoproduction of $C=+1$
(pseudo) scalar and tensor mesons $M$,  $\gamma p \to Mp' $
(with $p'$ either proton or its low-mass excitation), seems
to be a better signature for the odderon exchange
\cite{GIvod,Zhitnitsky}. At asymptotic energies, when the
$\rho$ and $\omega$ exchange contributions die out, such
processes will be dominated by the odderon exchange. These
reactions also have not yet been observed experimentally.

$\bullet$ The available calculations of the odderon
amplitude give only the first term --- "Born approximation"
--- of the reggeization program, which technically is carried
out by resummation of logarithms from loop corrections. It
is expected that, since the intercept of the odderon is
close to 1, the reggeized physical amplitude will be close
to the mentioned "Born" result. However, this approach, used
e.g. in refs.~\cite{dosh,hardpi}, requires care.

$\Box$ {\it The proton vertex.} In the widely used
quark-diquark model, the result depends strongly on the
clustering of quarks in the nucleon. With variation of this
clustering both the value of amplitude (by a factor $1\div
4$) and content of final state change strongly
\cite{Zakharov}. For example, the diagonal transition $p'=p$
is almost forbidden for point-like diquark (this particular
case is used in {\it the stochastic vacuum} calculation of
\cite{dosh}) and becomes not small (perhaps, dominant) for
more realistic finite diquark size.

The similar uncertainty takes place in the description of
dipion production via the hard odderon \cite{hardpi}. In the
pQCD approach the quark impact--factors for Pomeron and
odderon are identical \cite{GIvod}. However, the coupling of
Pomeron to the gluon content of proton can be essential, and
such coupling is absent for odderon \cite{IFGod}. It {\it
can} give different final $p'$ states for Pomeron and
odderon, reducing charge asymmetric interference.

$\Box$ {\it The second difficult point} is clearly seen in
the treatment of $\gamma p\to f_2p'$ process with its
helicity structure. Calculated "Born" term contains both
factorizable in helicity terms and non-factorizable terms.
Only factorizable terms should be regarded for the estimates
of cross sections under interest while non-factorizable
terms must be eliminated as having no relation to the
Reggeon (in our case --- the odderon).

The dominant $\gamma p\to f_2p'$ amplitude calculated in
ref.~\cite{dosh} is exactly non-factorizable: in
this amplitude spin flips in the vertices are correlated,
$\lambda_\gamma-\lambda_f = -(\lambda_p-\lambda_{p'}) =-1$,
and instead of dependence ${\cal A}\propto t$ following from
general property (\ref{bashel}), it does not vanish at
$t=0$. This non-factorizable term must be eliminated from
the result.

Therefore, two essential conclusions of ref.~\cite{dosh}
cannot be related to the odderon:\\({\it i}) the values of
cross sections estimated;\\ ({\it ii}) the predictions about
nucleon excitations dominance for the proton vertex.

$\bullet$ Recently, H1 collaboration tried to observe
odderon in the diffractive photoproduction. With event
selection including observation of only excited nucleon
states (required by \cite{dosh} results and hardly probable
in reality as it was mention above), the signal from odderon
was not found and the upper limits for various final states
were set. The previous discussion shows that the event
selection used in this experiment is misleading and
theoretical estimates of cross sections are irrelevant.

The calculations of hard odderon amplitudes were performed
in the lowest 3--gluon exchange approximation. The
calculations of ref.~\cite{GIvod} are valid only for very
large $|t|$ where cross sections are very small. The
calculations of refs.~\cite{oddqcd} consider
electroproduction at small $t$. The problem of reggeization
mentioned above in respect to \cite{dosh} was not considered
in these papers. The other uncertainties in these
calculations were discussed in sect.~\ref{sechard}.

$\blacksquare$ Therefore, despite the fact that the odderon
is a necessary feature of QCD motivated description of
diffractive type processes, up to the moment we have no
approach giving reliable estimations for the processes like
$\gamma p\to f_2p'$, etc. at $k_\bot\lesssim 1$ GeV. That is
the reason why in ref.~\cite{GIN1} and in the text above we
use completely phenomenological description and present only
calculations for the lowest total odderon cross section
exceeding effect of other mechanisms. \\

I am thankful to D. Anipko, S. Brodsky, A. Denig, A.V.
Efremov, V.A.~Ilyin, D.Yu.~Iva\-nov, I.P.~Ivanov,
K.~Hencken, S.~Kolb, B.~Kopeliovich, N.N.~Nikolaev, A.~Pak,
O.~Panella, B.~Pire, M.~Ryskin, V.~Saveliev, A.~Schiller,
V.G.~Serbo, L.~Szymanowski, O.V.~Teryaev, H.~Young for the
discussions of different problems mentioned in this work.
This paper is supported by grants RFBR 02-02-17884, RFBR
00-15-96691, INTAS 00-00679 and grant 015.02.01.16 Russian
Universities. I am also grateful to Landau--Centro Volta
Network, O.~Panella and J.~Srivastava for the warm
hospitality during staying in INFN Perugia where this paper
was completed.

\end{document}